\documentclass{JHEP}
\usepackage{graphicx}
\usepackage{epsfig}
\usepackage{amssymb}
\usepackage{amsmath}
\usepackage{float}

\newcommand{\bea}{\begin{eqnarray}}
\newcommand{\eea}{\end{eqnarray}}
\newcommand{\bean}{\begin{eqnarray*}}
\newcommand{\eean}{\end{eqnarray*}}
\newcommand{\nn}{\nonumber \\}

\def\W #1{\widetilde{#1}}
\def\WH #1{\widehat{#1}}

\def\Tr{\mathop{\rm Tr}}

\def\eref#1{(\ref{#1})}

\def\a{{\alpha}}

\def\b{{\beta}}

\def\Sl{\sum\limits}
\def\Label#1{\label{#1}%
  \smash{\hbox to0pt{\raise1ex\hbox{\tiny[#1]}\hss}}}

\allowdisplaybreaks
\title{Dual-color decompositions at one-loop level in Yang-Mills theory}

\author{Yi-Jian Du ${}^{a,b}$, Bo Feng ${}^{c,d}$, Chih-Hao Fu ${}^e$
~~~~~~~~~~~~~~\\${}^a$ Department of Physics and Center for Field Theory
and Particle
 Physics, Fudan University, Shanghai 200433, P.R China\\
 ${}^b$ Department of Astronomy and Theoretical Physics, Lund University
SE-22362 Lund, Sweden\\
 ${}^c$ Zhejiang Institute of Modern Physics, Zhejiang
University, 38 Zheda Road Hangzhou, 310027 P.R China\\${}^d$
Center of Mathematical Science, Zhejiang
University, 38 Zheda Road Hangzhou, 310027 P.R China\\
${}^e$ Department of Electrophysics, National Chiao Tung University\\
1001 University Road, Hsinchu, Taiwan, R.O.C.\\
~~~~~~~\\
 \email{ yjdu@fudan.edu.cn; b.feng@cms.zju.edu.cn; zhihaofu@nctu.edu.tw \hskip0.5cm} }
\preprint{LU-TP 14-07}
\date{\today}
\abstract{In this work, we extend the construction of  dual color decomposition in Yang-Mills theory to one-loop level, i.e.,
we show how to write  one-loop integrands in Yang-Mills theory to the dual DDM-form and the dual trace-form.
In dual forms, integrands are decomposed in terms of color-ordered one-loop integrands for color scalar theory with proper dual color coefficients.
In dual DDM decomposition,
The dual color coefficients can be obtained directly from BCJ-form by applying Jacobi-like identities for kinematic factors.
In dual trace decomposition, the dual trace factors can be obtained by imposing one-loop KK relations, reflection relation and their relation with the kinematic factors in dual DDM-form.
}


\begin{document}
%%%%%%%%%%%%%%%%%%%%%%%%%
\section{Introduction}
%%%%%%%%%%%%%%%%%%%%%%%%%%
One of the significant recent progresses in the study of scattering amplitude is the discovery of color-kinematic duality made by Bern, Carrasco and Johansson (BCJ) \cite{Bern:2008qj}. It is conjectured  that
 a generic $L$-loop Yang-Mills amplitude can be written to the double-copy formula (we will call the
 form as BCJ-form)
\bea
\mathcal{A}_{tot}^{L}=i^Lg^{m-2+2L}\Sl_{D_i}\int\prod\limits_{j=1}^l{d^Dl_j\over (2\pi)^D}{1\over S_i}{n_i(\emph{l})c_i\over \prod_k P_{ki}(\emph{l})},~~\Label{L-loop double copy}
\eea
where the sum is  over all  possible cubic Feynman-like diagrams and the $S_i$ is
the symmetric factor.
In the formula, the $c_i$ is the color factor given by the group structure constants $f^{abc}$
and $n_i(\emph{l})$ is the kinematic factor satisfying the following properties:
whenever two color factors  $c_i, c_j$ are related by antisymmetry or
three color factors $c_i, c_j, c_k$ by Jacobi identity, so are
the corresponding kinematic factors $n$.
\bea
{\rm antisymmetry}:& c_i=-c_j & \Rightarrow n_i=-n_j\nn
{\rm Jacobi-like~identity}:& c_i+c_j+c_k=0 & \Rightarrow n_i+n_j+n_k=0.~~\label{color-kinematic-duality}
\eea
 The BCJ-form of Yang-Mills theory was proved at the tree-level
from string theory in \cite{BjerrumBohr:2009rd,Stieberger:2009hq}, from  twistor string
theory in \cite{Cachazo:2012uq, Cachazo:2013gna, Cachazo:2013hca, Cachazo:2013iea}, from field theory
in \cite{Feng:2010my, Jia:2010nz, Chen:2011jxa} and further studied in \cite{Sondergaard:2009za, Tye:2010dd, BjerrumBohr:2010zs, Tye:2010kg, Mafra:2011kj, Monteiro:2011pc, BjerrumBohr:2012mg,Fu:2012uy, Monteiro:2013rya} (see also a nice review \cite{Sondergaard:2011iv}).
The BCJ-form at the loop level is still a conjecture, but many studies have appeared\cite{BjerrumBohr:2011xe,Boels:2011tp,Boels:2011mn, Du:2012mt,Boels:2012ew, Carrasco:2012ca,Carrasco:2011mn, Bjerrum-Bohr:2013iza, Bern:2013yya, Boels:2013bi, Nohle:2013bfa}. The apparent equal-footing treatment on the color and kinematic factors of \eref{L-loop double copy}
introduced a very interesting perspective to the understanding of the
structure of Yang-Mills amplitudes. To see its implication more clearly, let us
review some results at  tree-level.

%%%%%%%%%%%%%%%%%%%%%%
\subsection*{The tree-level case}
%%%%%%%%%%%%%%%%%%%%%%
Based on results \cite{Kleiss:1988ne, DelDuca:1999rs} as well as \cite{Bern:2008qj}, at tree-level we can write Yang-Mills amplitudes  in the following three color decomposition forms:
\bea {\rm double-copy \, (or \, BCJ-) \, form}: &~~~ & {\cal A}_{tot} = \sum_i { c_i n_i\over
D_i}~~~\Label{BCJ-form}\\
{\rm Trace~form}: &~~~ &{\cal A}_{tot}  = \sum_{\sigma\in
S_{n-1}}{\rm Tr} (T^{\sigma_1}... T^{\sigma_n})
A(\sigma)~~~\Label{Trace-form}\\
{\rm DDM~form}: &~~~ & {\cal A}_{tot}  =  \sum_{ \sigma\in S_{n-2}}
c_{1|\sigma(2,..,n-1)|n} A(1,\sigma,n)~~~\Label{DDM-form} \eea
Here Roman $A$s represent color ordered amplitudes, $T^a$
are generators of $U(N)$ in fundamental representation
%is the matrix of
%fundamental representation of $U(N)$ group
and
$c_i,c_{1|\sigma(2,..,n-1)|n} $
%are constructed using the structure
represent strings of  structure
constants $f^{abc}$
\bea c_{1|\sigma(2,..,n-1)|n}=f^{1 \sigma_2 x_1} f^{x_1
\sigma_3 x_2}... f^{x_{n-3} \sigma_{n-1} n}~.~~\Label{DDM-c}\eea
 Among these three forms, the relation between trace-form and DDM-form %is
 has been
well understood %by
using the following two properties of the Lie algebra of $U(N)$ gauge group.
(See  Ref. \cite{DelDuca:1999rs})
\bea {\rm Property~One:} & ~~~ & (f^a)_{ij}= f^{aij}={\rm Tr}( T^a[
T^i,
T^j]),~~~\Label{group-1}\\
{\rm Property~Two:} & ~~~ &  \sum_a {\rm Tr}( X T^a) {\rm Tr}(T^a
Y)= {\rm Tr}( XY)~~\Label{group-2}\eea
On the other hand, the existence of BCJ-form \eref{BCJ-form} is very nontrivial and recently many works
have been devoted to its understanding as reviewed in the previous paragraph.
 For special helicity
configurations, it was shown that the kinematic numerators correspond to
area-preserving diffeomorphism algebra\cite{Monteiro:2011pc,BjerrumBohr:2012mg}. Using this idea,
 an explicit construction
of the BCJ-numerators $n_i(\emph{l})$ was given in \cite{Fu:2012uy}, thereby providing a support to  an algebra-manifest formulation.
%%Using Berkovits' pure spinor formalism
Also that Mafra, Schlotterer and Stieberger have given an explicit construction in \cite{Mafra:2011kj}
using Berkovits' pure spinor formalism.
Finally, using the twistor string theory, Cachazo, He and Yuan gave an algorithm for
$n_i(\emph{l})$ using solutions from
scattering equations  \cite{Cachazo:2013iea}.

Although it is very hard\footnote{In fact, we are not clear how to do so.} to derive  the BCJ-form
from the trace-form or the DDM-form, it is not hard to establish the  trace-form and the DDM-form from the BCJ-form.
Explicitly, using  Jacobi relations,  one  can  construct a basis for all color factors $c_i$,
which are nothing, but the factors given in \eref{DDM-c}. Knowing the basis,  we can expand arbitrary
color factor  $c_i=\sum_\sigma \a_\sigma c_{1|\sigma |n}$ and putting them back to \eref{BCJ-form}. After collecting terms according to factors $c_{1|\sigma |n}$, \eref{BCJ-form} becomes
 the following form
\bea {\cal A}_{tot}  =  \sum_{ \sigma\in S_{n-2}}
c_{1|\sigma(2,..,n-1)|n}\WH  A(1,\sigma,n)~~~~\Label{DDM-form-tilde}\eea
Then we need to  ask  whether the $  \WH  A(1,\sigma,n)$ defined here is the same  as $    A(1,\sigma,n)$
given in \eref{DDM-form}. This identification can been done either from $c_{1|\sigma(2,..,n-1)|n}$ as the basis of color factor and both $\WH A$ and $A$ are minimum gauge invariant objects, or using the KLT relation \cite{KLT}
in \cite{Du:2011js}.

%%%%%%%% pause

However, since in  the double-copy formulation (BCJ-form) $n_i$  acquire the same status as $c_i$, it is natural to exchange the roles between $c_i$ and $n_i$ and consider the following two dual
formulations
\bea {\rm Dual~Trace-form}: &~~~ &{\cal A}_{tot}  = \sum_{\sigma\in
S_{n-1}} \tau_{\sigma_1... \sigma_n}
\W A(\sigma)~~~\Label{Dual-Trace-form}\\
{\rm Dual~DDM-form}: &~~~ & {\cal A}_{tot}  =  \sum_{ \sigma\in
S_{n-2}} n_{1|\sigma(2,..,n-1)|n}\W
A(1,\sigma,n)~~~\Label{dual-DDM-form} \eea
where $\W A$ is the color ordered scalar amplitude with $f^{abc}$ as its cubic
coupling constant (See  Ref.   \cite{Bern:1999bx,
Du:2011js}) and $\tau$ is required to be cyclic invariant.

The idea of  dual DDM-form first appeared in the literature   in \cite{Bern:2010yg}.
Using  Jacobi identity we can find the basis $n_{1|\sigma |n}$ and expand any other
 $n_i=\sum_\sigma n_{1|\sigma |n}$ as we did for color factor $c_i$. Putting them back to
 BCJ-form \eref{BCJ-form} and collecting terms  leads to the form given in \eref{dual-DDM-form}. However, it
is not clear whether the $\W  A$ obtained by this way is  the same color-ordered scalar amplitude with $f^{abc}$
coupling as claimed in \eref{dual-DDM-form}.
To establish this fact, one idea is to use the KLT relation as was done in \cite{Du:2011js}
to establish the existence of dual DDM-form ( however, now we need to use the off-shell BCJ relation for gauge theory amplitudes presented in \cite{Fu:2012uy}). Then by the independence of basis $n_{1|\sigma |n}$,
the identification of $\W A$ is done.

Based on the established  existence of dual DDM-form, the dual trace-form was conjectured in \cite{Bern:2011ia} with explicit constructions given for the first few lower-point amplitudes.
In addition two  constructions for the dual trace-form was discussed in \cite{Du:2013sha} and
\cite{Fu:2013qna}.

%\end{itemize}
%

%%%%%%%%%%%%%%%%%%%%%%%%
\subsection*{One-loop level}
%%%%%%%%%%%%%%%%%%%%%%%%%
%%
Having reviewed the tree-level case, now we move to the one-loop case,
which will be our focus in this paper. At one-loop level we have the following three   color
decompositions:
\bea {\rm BCJ-form}: && {\cal A}^{1-loop}_{tot} = ig^{n}\Sl_{diagrams~\Gamma^i}\int\prod {d^D \emph{l}\over (2\pi)^D}{1\over S_i}{n_i(\emph{l})c_i\over \prod_k P_{ki}(\emph{l})}~~~\Label{1loop-BCJ-form}\\
{\rm Trace-form}: &&{\cal A}^{1-loop}_{tot}  = N_c\sum_{S_n/Z_n}{\rm Tr} (T^{\sigma_1}... T^{\sigma_n})
A_{n;0}(\sigma_1,...,\sigma_n)\nn & & +\sum_{m=1}^{\lfloor n/2\rfloor}\Sl_{\sigma\in S_n/S_{n;m}} {\rm Tr}( T^{\sigma_1}...T^{\sigma_{n-m}}){\rm Tr}( T^{\sigma_{n-m+1}}...T^{\sigma_{n}})
A_{n-m;m}(\sigma_1,...,\sigma_{n-m};\sigma_{n-m+1},...,\sigma_{n})\nn
&&~~~~~~~~~~~~~~~~~~~~~~~~~~~~~~~~~~~~~~~~~~~~~~~~~~~~~~~~~~~~~~~~~~~~~~~~~~~~~~~~~~~~~~~~~~\Label{1loop-Trace-form}\\
{\rm DDM-form}: && {\cal A}^{1-loop}_{tot}  =  \sum_{ \sigma\in S_{n-1}/R}
f^{ x_n\sigma_1 x_1}f^{x_1 \sigma_2 x_2}....f^{x_{n-1} \sigma_n x_n} A_{n,0}(\sigma_1,...,\sigma_n)~~~\Label{1loop-DDM-form} \eea
In \eqref{1loop-Trace-form}, $Z_n$ denote cyclic symmetry and $S_{n;m}$ is the subsets of $S_n$ that leaves the double-trace structure invariant, $\lfloor n/2\rfloor$ is the greatest integer less than or equal to $n/2$. In \eqref{1loop-Trace-form}, $R$ denotes reflection.
Among these three forms, the last two, i.e., \eref{1loop-Trace-form} and \eref{1loop-DDM-form}, are well
established while the first one \eref{1loop-BCJ-form} is still a conjecture.

The trace-form \eref{1loop-Trace-form} was given by Bern, Dixon, Dunbar and Kosower in
\cite{Bern:1994zx}. In the formula there are single and double trace parts, %. However,
where the partial amplitude $A_{n-m;m}(\sigma_1,...,\sigma_{n-m};\sigma_{n-m+1},...,\sigma_{n})$ associated to the
double trace part can be obtained by linear combination of those to the single trace part $A_{n;0}(\sigma_1,...,\sigma_n)$. In other words, we do not need to calculate $A_{n-m;m}(\sigma_1,...,\sigma_{n-m};\sigma_{n-m+1},...,\sigma_{n})$ for one-loop amplitudes. The DDM-form, on the other hand, was
given in \cite{DelDuca:1999rs}, where the sum is over noncylic permutation up to reflections $R: (12...n)=(n...21)$. Here the $A_{n;0}(\sigma_1,...,\sigma_n)$ in \eref{1loop-DDM-form}
is nothing but the single trace partial amplitude appeared in \eref{1loop-Trace-form}.
In fact, starting from DDM-form, it is easy to derive the trace-form as demonstrated in \cite{DelDuca:1999rs}.
As a byproduct, the relation between single and double trace partial amplitudes will appear automatically.

Assuming the existence of \eref{1loop-BCJ-form},  to go from BCJ-form to
trace-form \eref{1loop-Trace-form} and DDM-form \eref{1loop-DDM-form} is easy.
As shown by \cite{DelDuca:1999rs}, for one-loop color factors,  annuli of structure constants of the form
\bea c^{1-loop}(\sigma_1...\sigma_n)\equiv f^{ x_n\sigma_1 x_1}f^{x_1 \sigma_2 x_2}....f^{x_{n-1} \sigma_n x_n} ~~~~\Label{1loop-color-basis}\eea
serve as a basis. Using it we can expand any $c_i$ in BCJ-form and collect terms with factor $c^{1-loop}(\sigma_1...\sigma_n)$. These terms as a whole can be denoted by $\WH A_{n,0}(\sigma_1,...,\sigma_n)$.
Again the problem is whether  it is equal to the one  $ A_{n,0}(\sigma_1,...,\sigma_n)$ defined in \eref{1loop-DDM-form}? The identification is again easy by using the following facts:
 the color basis $c^{1-loop}(\sigma_1...\sigma_n)$ is independent to each
other, and $\WH A$ and $A$ are gauge invariant objects.

%\end{itemize}
%
%%%%%%%%%%%%%%%%%%%%%%%%%%%%%%%%%%%%%%%%%%%%%%%%%%%%%%%%%%
%%%%%%%%%%%%%%%%%%%%%%%%%%%%%%%%%%%%%%%%%%%%%%%%%%%%%%%%%%
%%%%%%%%%%%%%%%%%%%%%%%%%%%%%%%%%%%%%%%%%%%%%%%%%%%%%%%%%%
%%%%%%%%%%%%%%%%%%%%%%%%%%%%%%%%%%%%%%%%%%%%%%%%%%%%%%%%%%
%%%%%%%%%%%%%%%%%%%%%%%%%%%%%%%%%%%%%%%%%%%%%%%%%%%%%%%%%%
%%%%%%%%%%%%%%%%%%%%%%%%%%%%%%%%%%%%%%%%%%%%%%%%%%%%%%%%%%

The above discussions are parallel to the one given for tree-level case. Considering the
duality between $n_i$ and $c_i$ in \eref{1loop-BCJ-form}, it is natural to investigate
 the dual form \cite{Bern:2011ia},  where the interest of  this paper lies\footnote{In supergravity theory, a DDM-form of decomposition at one-loop level of supergravity has already been suggested in \cite{Bern:2011rj} and the dual DDM-form in Yang-Mills theory has the similar form. Nevertheless, in this work, we would like to give a general discussion on dual DDM-form at one-loop level in Yang-Mills theory in the introduction and some explicit examples in section \ref{sec:dual-ddm}, because the dual DDM-form is crucial for the construction of dual trace-form. }. Unlike color numerator,
the $n_i$ depends on the loop momentum in general, so  the dual formulations at one-loop should
be given by\footnote{One may notice that in one-loop DDM form \eqref{1loop-DDM-form}, reflection has been modded out, in the dual DDM-form, we just leave the reflection symmetry and only consider it when we discuss on the dual trace-form.  }
\bea
{\cal A}_{1-loop}=ig^{n}\int {d^D \emph{l}\over (2\pi)^D}\Sl_{\sigma\in S_{n-1}} n^{DDM}_{1,\sigma}(\emph{l})\W I(1,\sigma).~~\Label{1loop-Dual-DDM}
\eea
for the dual DDM-form, and
\bea
{\cal A}_{1-loop}=ig^{n}\int {d^D\emph{ l}\over (2\pi)^D}\sum_{m=0}^{\lfloor n/2\rfloor}\Sl_{\sigma\in S_n/S_{n;m}}\tau_{\{\sigma_1,\dots,\sigma_{n-m}\},\{\sigma_{n-m+1},\dots,\sigma_n\}}(\emph{l})\W I(\{\sigma_1,\dots,\sigma_{n-m}\};\{\sigma_{n-m+1},\dots,\sigma_n\}),~~\Label{Dual-trace}\nn
\eea
for the dual trace-form\footnote{Since we only discuss on one-loop case, we will use $n$ and $\tau$ and $\W I$ instead of $n^{1-loop}$, $\tau^{1-loop}$, $\W I^{1-loop}$ for convenience.}. In other words, it is the {\sl integrand} taking the dual form.
Using the same idea, to get dual DDM-form from BCJ-form, first we  need to find a suitable basis for $n_i$, then put it back to BCJ-form and collect terms to get the dual DDM-form. After this step, again the key step is to identify
what integrand these collected terms correspond to. An intuition might be provided by
making use of the existing tree-level results. Naively if we start with an $(n+2)$-point Feynman diagrams, a
one-loop diagram can be constructed by connecting the $(n+1)$-th and $(n+2)$-th external lines. Since we know
%%%the tree-level amplitude is the one of color-ordered scalar theory,
at tree-level these collected terms correspond to color-ordered scalar amplitudes,
it is very natural to do the identification
at the one-loop level.

Now we make above observation more accurately.  Starting from BCJ-form, note that in order to
 get DDM-form, one decomposes color numerators $c_i=\sum_\a \kappa_{i\a} c_{\a}$, where $c_\a$ is
a basis of the color part constructed  only by antisymmetry and the Jacobi identity. After that we get
\bea \sum_\a c_\a \left( \sum_{i:diagrams} \kappa_{i\a} { n_i\over D_i}\right)~~~\Label{test-1}\eea
where the   factors inside the bracket  together constitute the color-ordered gauge theory amplitude.
Likewise, to get the dual DDM-form  one decomposes kinematic numerators according to
 $n_i=\sum_\a \kappa_{i\a} n_{\a}$ where $n_\a$ is another basis of the kinematic part constructed, again,   only by antisymmetry and the Jacobi identity. It is crucial the construction of basis used only the topology of cubic
 diagrams, thus we can take  bases $c_\a$ and $n_\a$ sharing the same diagram topology. In other words,
we should have the same expansion coefficients $\kappa_{i\a}$ for both constructions of DDM-form and dual DDM-form
\bea \sum_\a n_\a \left( \sum_{i:diagrams} \kappa_{i\a} { c_i\over D_i}\right)~~~~\Label{test-2}\eea
Comparing \eref{test-1} and \eref{test-2}, we see that the difference is just the exchange
of $c_i\leftrightarrow n_i$, whereas $c_i$ and $n_i$ correspond to the same cubic diagram.
Because  the $c_i$ has local construction, i.e., each cubic vertex is decorated with the coupling constant $f^{abc}$, we conclude that the integrand of one-loop dual DDM-form is indeed the one of  color-ordered scalar theory. In section \ref{sec:dual-ddm}, we will use explicit calculations to demonstrate above arguments.

Having obtained the dual DDM-form, the next step is to construct the dual trace-form.
 Going from dual DDM-form  to dual trace-form, we need to find a way to rewrite
 basis kinematic numerators $n_\a$ to a linear combination of some kind "single trace" part
 and "double trace" part $n_{trace}$ as did for tree-level case in  \cite{Bern:2011ia, Du:2013sha, Fu:2013qna}.
 However note that, as in the tree-level case, the number of $n_{trace}$ is much more than $n_\a$, thus
proper extra relations need to be manually imposed in order to solve $n_{trace}$ by $n_\a$. Choosing the appropriate relations is nevertheless far from trivial, in particular one needs to avoid over-constraint and maintain relabeling symmetry if possible. A $4$-point example  at one-loop level was provided by Bern and Dennen in \cite{Bern:2011ia}, where cyclic and KK-relations were implemented. In section \ref{sec:dual-trace},
we will generalize the result in \cite{Bern:2011ia} and give a general algorithm for the construction of
dual trace-form at one-loop. Our algorithm gives  the solution satisfying natural relabeling symmetry.

In section 4, we will use the relabeling symmetry to give another construction of dual trace-form. Finally
in section \ref{sec:conclusion} a brief conclusion is given.

%%%%%%%%%%%%%%%%%%%%%%%%%%%%%%%
\section{Dual DDM-form}
%%%%%%%%%%%%%%%%%%%%%%%%%%%%%%%
\label{sec:dual-ddm}

Having the general discussion for the dual DDM-form at one-loop, in this section, we will use explicit example
to demonstrate the construction. In the  discussion below we will follow the convention
where the loop momentum $\emph{l}$ is defined to be the momentum carried by the propagator next to  leg $1$.
%%%%%%%%%%%%%%%%%%%%%%%%%%%
\subsection{Two-point example}
%%%%%%%%%%%%%%%%%%%%%%%%%%%
%%
\begin{figure}[h!]
  \centering
 \includegraphics[width=0.6\textwidth]{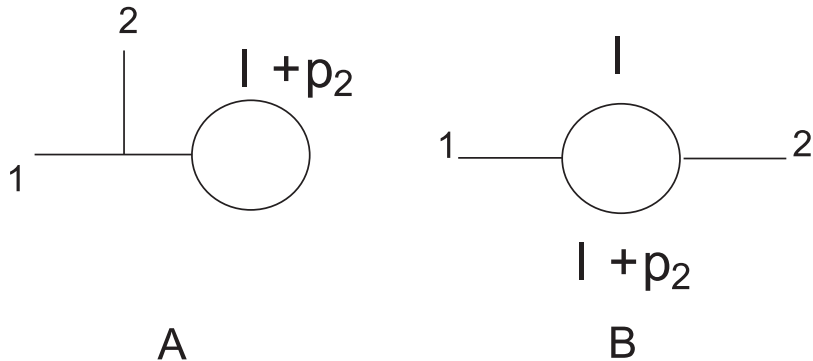}
 \caption{Feynman-like diagrams for two-point one-loop integrand using only cubic vertex.}\label{2-pt}
\end{figure}
%%%
%%%
For two point case, using only cubic vertex, two diagrams A and B are constructed as given in Figure \ref{2-pt}. With our convention, the corresponding integrands are
\bea
I_{A}(l)={C_An_A(l)\over s_{12}(l+p_1+p_2)^2},~~~~ I_B(l)={C_{B}n_{B}(l)\over l^2(l+p_2)^2}.
\eea
where
\bea
C_A=f^{12e}f^{ee'e'},~~~~ C_B=f^{e1e'}f^{e'2e}
\eea
and the Einstein summation convention has been used. Since the structure constant is antisymmetric, $f^{ee'e'}=0$, so $C_A=0$. The two-point one-loop integrand becomes
\bea
I_{2-pt}(l)=I_B(l)=n_{B}(l)\left[{f^{e1e'}f^{e'2e}\over l^2(l+p_2)^2}\right].
\eea
Comparing with \eref{1loop-Dual-DDM}, we see that the part inside the bracket in above equation is nothing, but
the integrand $\W I(1,2)$ we are looking for. It is obvious from the expression that $\W I(1,2)$ is the one-loop integrand of color-ordered scalar theory with two external lines.

%%%%%%%%%%%%%%%%%%%%%%%%%%%%%
\subsection{Three-point example}
%%%%%%%%%%%%%%%%%%%%%%%%%%%%%%
\begin{figure}[h!]
  \centering
 \includegraphics[width=0.7\textwidth]{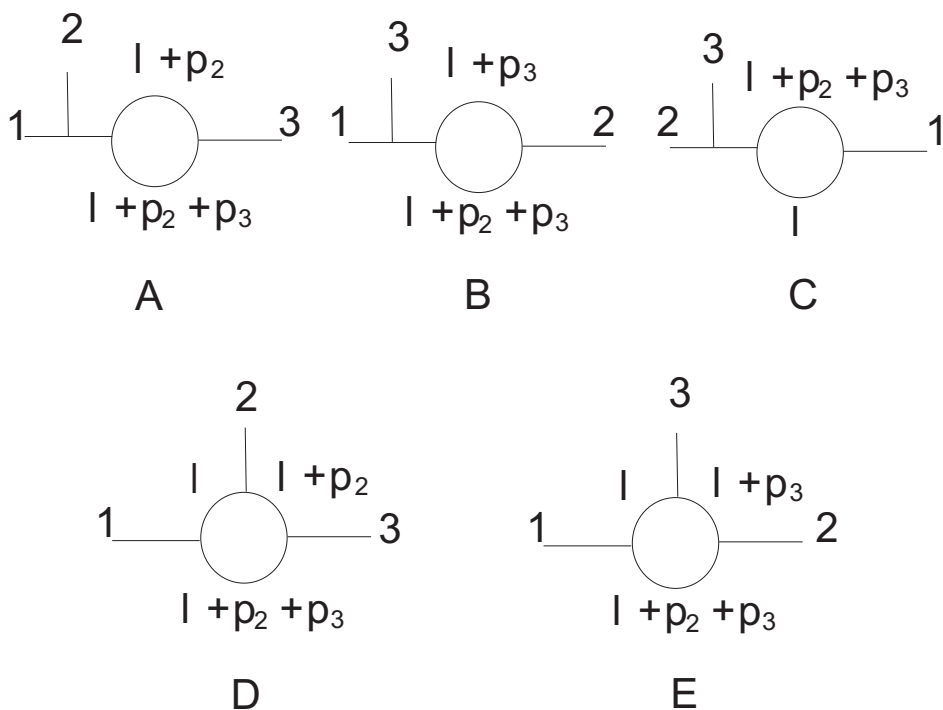}
 \caption{Feynman-like diagrams for three-point one-loop integrand.}\label{3-pt}
\end{figure}
For the three-point case, using only cubic vertex there are three kinds of topologies (see Figure \ref{3-pt}).
The first kind of topologies is the tadpole diagrams, i.e., there is only one line connected to the loop. Because
the antisymmetric property of group structure constants, the contribution is zero, just like the diagram A of
two-point case in  Figure \ref{2-pt}. The second kind of topologies has two lines connected to the loop directly, i.e.,  diagrams  A, B, C in Fig. \ref{3-pt}. Expressions are given by
\bea
I_A(l)={C_An_A(l)\over s_{12}(l+p_2)^2(l+p_2+p_3)^2}, I_B(l)={C_An_B(l)\over s_{13}(l+p_3)^2(l+p_3+p_2)^2},I_C(l)={C_C n_C(l)\over s_{23} l^2(l+p_2+p_3)^2}.
\eea
where color factors $C_A, C_B, C_C$ can easily be read out from corresponding diagrams.
The third kind of topologies have  three lines connected to the loop directly, i.e.,  diagrams D, E in Fig. \ref{3-pt} and expressions are
\bea
I_D(l)={C_Dn_D(l)\over l^2(l+p_2)^2(l+p_2+p_3)^2}, ~~~I_E(l)={C_En_E(l)\over l^2(l+p_3)^2(l+p_3+p_2)^2}.
\eea
where $C_D, C_E$ are corresponding color factors.
 Using Jacobi-like identity and taking $n_D$ and $n_E$ as basis,  we find following expansions
\bea
n_A(l)=n_D(l)-n_E(l+p_2),~~ n_B(l)=n_E(l)-n_D(l+p_3),~~n_C(l)=n_D(l)-n_E(l).
\eea
Thus $I_A$, $I_B$, $I_C$ can be written as
\bea
I_A(l)&=&{C_A[n_D(l)-n_E(l+p_2)]\over s_{12}(l+p_2)^2(l+p_2+p_3)^2}={C_An_D(l)\over s_{12}(l+p_2)^2(l+p_2+p_3)^2}-{C_An_E(l)\over s_{12}l^2(l+p_3)^2}+T_A\nn
I_B(l)&=&{C_B[n_E(l)-n_D(l+p_3)]\over s_{13}(l+p_3)^2(l+p_3+p_2)^2}={C_Bn_E(l)\over s_{13}(l+p_3)^2(l+p_3+p_2)^2}-{C_Bn_D(l)\over s_{13}(l)^2(l+p_2)^2}+T_B\nn
I_C(l)&=&{C_C[n_D(l)-n_E(l)]\over s_{23}l^2(l+p_2+p_3)^2},
\eea
where $T_A$ and $T_B$ are terms integrated to zero\footnote{It can easily be seen by shifting the loop momentum of
the first term. }
\bea
T_A&=&{C_A\over s_{12}}\left[{n_E(l)\over l^2(l+p_3)^2}-{n_E(l+p_2)\over (l+p_2)^2(l+p_2+p_3)^2}\right],\nn
T_B&=&{C_B\over s_{13}}\left[{n_D(l)\over l^2(l+p_2)^2}-{n_D(l+p_3)\over (l+p_3)^2(l+p_3+p_2)^2}\right].
\eea
Up to terms integrated to zero, the total integrand is given as
\bea
& & I(1,2,3)(l)=n_D(l)\Biggl[{C_A\over s_{12}(l+p_2)^2(l+p_2+p_3)^2}-{C_B\over s_{13}(l)^2(l+p_2)^2}+{C_C\over s_{23} l^2(l+p_2+p_3)^2}\nn
&&+{C_D\over l^2(l+p_2)^2(l+p_2+p_3)^2}\Biggr]+n_E(l)\Biggl[-{C_A\over s_{12}l^2(l+p_3)^2}+{C_B\over s_{13}(l+p_3)^2(l+p_3+p_2)^2}-{C_C\over s_{23} l^2(l+p_2+p_3)^2}\nn
&&+{C_E\over l^2(l+p_3)^2(l+p_3+p_2)^2}\Biggr]\equiv n_D(l)\W I(1,2,3)(l)+n_E(l)\W I(1,3,2)(l),
\eea
Above form is exactly the dual DDM-form for one-loop amplitude \eref{1loop-Dual-DDM}, where
 $n_D$ and $n_E$ are just the kinematic basis. The corresponding expressions for $\W I(1,2,3)(l)$
 and $\W I(1,3,2)(l)$ are again the three-point one-loop integrands of color-ordered scalar theory.

%%%%%%%%%%%%%%%%%%%%%
\subsection{Four-point example}
%%%%%%%%%%%%%%%%%%%%%%
\begin{figure}[h!]
  \centering
 \includegraphics[width=0.7\textwidth]{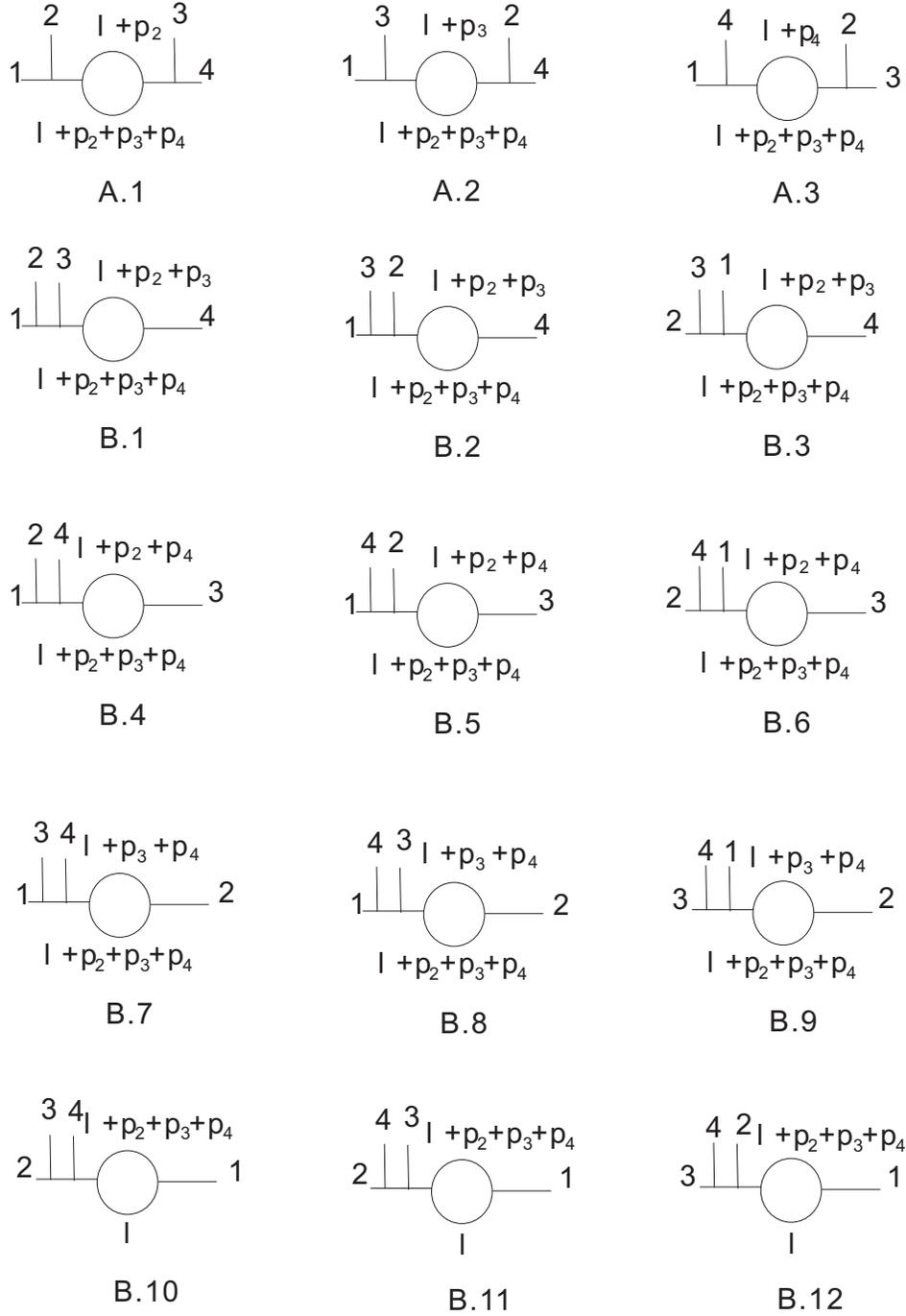}
 \caption{Feynman-like diagrams with two lines connected to the loop in four-point case.}\label{4-pt-1}
\end{figure}
\begin{figure}[h!]
  \centering
 \includegraphics[width=0.7\textwidth]{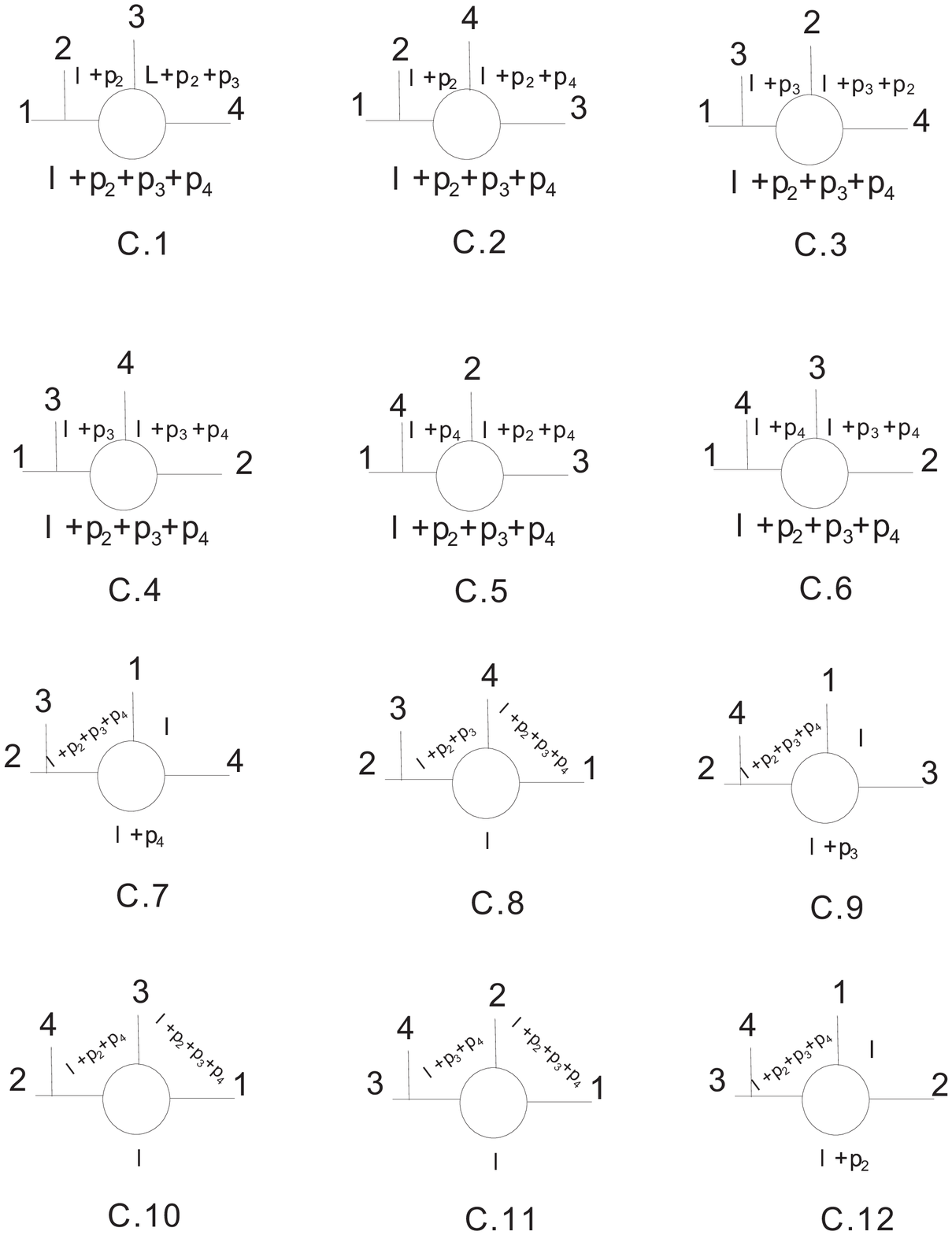}
 \caption{Feynman-like diagrams with three lines connected to the loop in four-point case.}\label{4-pt-2}
\end{figure}
\begin{figure}[h!]
  \centering
 \includegraphics[width=0.7\textwidth]{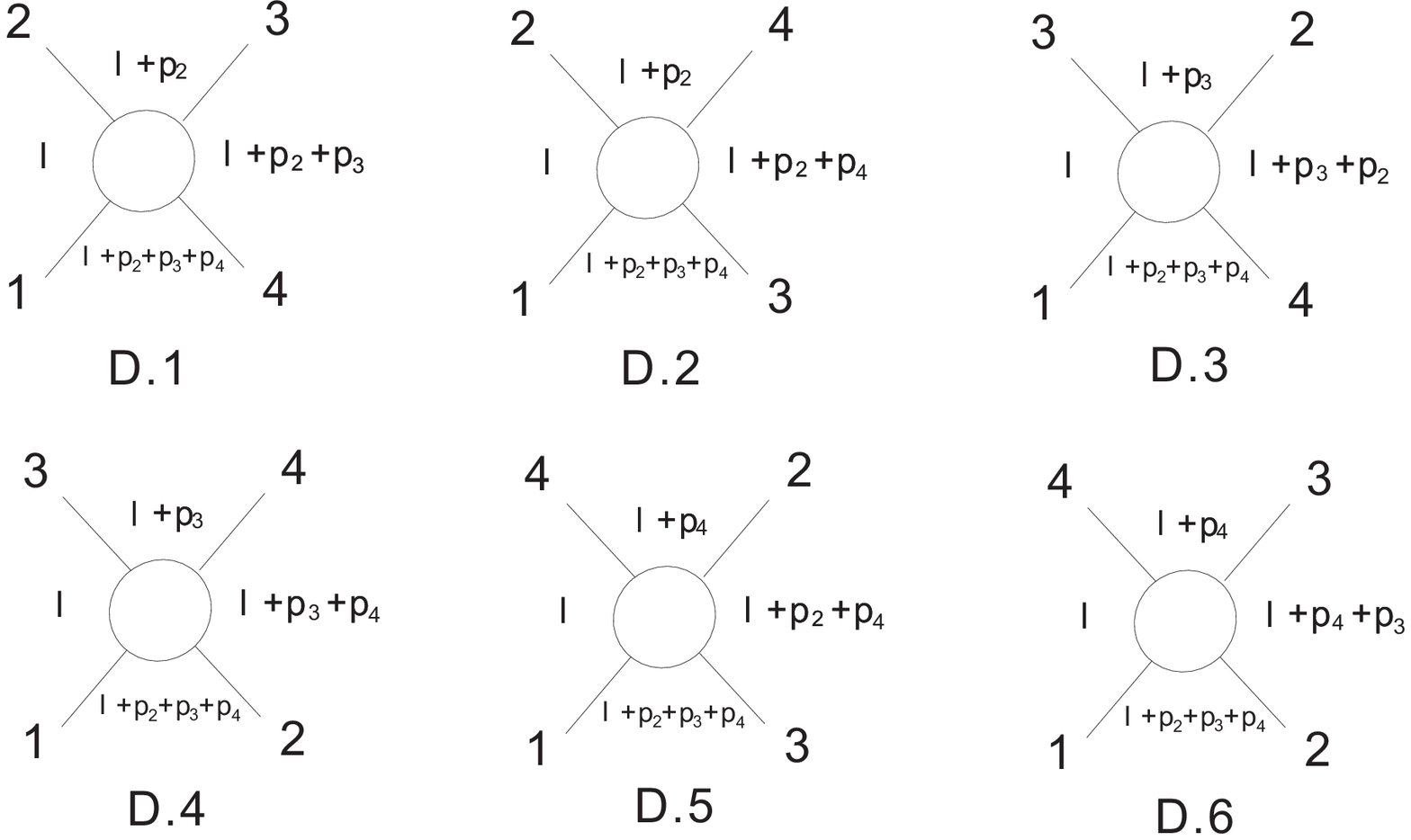}
 \caption{Feynman-like diagrams with four lines connected to the loop in four-point case.}\label{4-pt-3}
\end{figure}
For four-point case, there are many diagrams and they  can be classified as follows:
\begin{itemize}
\item (1) For tadpole diagrams with  only one line connected to the loop directly, their contributions are zero due to the antisymmetry of group structure constant.

\item (2) For diagrams with two lines connected to the loop directly, they are listed in Fig. \ref{4-pt-1}.
Their corresponding expressions can be read out easily.

\item (3) For diagrams with three lines connected to the loop directly, they are listed in Fig. \ref{4-pt-2}.
From these diagrams, it is easy to write down corresponding expressions.

\item (4) For diagrams with four lines connected to the loop directly, they are listed in Fig. \ref{4-pt-3}. From these diagrams, it is easy to write down corresponding expressions.

\end{itemize}
We will choose the kinematic basis $n_\a$ as these given by Fig. \ref{4-pt-3}( i.e., the  D.1-D.6) in dual DDM-form, and expand other $n_i$ given by Fig. \ref{4-pt-1} and Fig. \ref{4-pt-2} using Jacobi identities.
For example, the coefficient of  $n^{DDM}_{1234}$ will get contributions from diagrams A.1, A.3, B.1, B.3, B.5, B.8, B.9, B.10, B.12, C.1, C.5, C.8, C.12, D.1, as
 \bea
 &&{C_{A.1}\over s_{12}s_{34}(l+p_2)^2(l-p_1)^2}+{C_{A.3}\over s_{14}s_{23}(l+p_4)^2(l-p_1)^2}\nn
 &+&{C_{B.1}\over s_{12}s_{123}(l+p_2+p_3)^2(l-p_1)^2}+{C_{B.3}\over s_{23}s_{123}(l+p_2+p_3)^2(l-p_1)^2}+{C_{B.5}\over s_{14}s_{124}(l+p_2+p_4)^2(l-p_1)^2}\nn
 &+&{C_{B.7}\over s_{13}s_{134}(l+p_3+p_4)^2(l-p_1)^2}+{C_{B.9}\over s_{34}s_{134}(l+p_3+p_4)^2(l-p_1)^2}+{C_{B.10}\over s_{23}s_{234}l^2(l-p_1)^2}\nn
 &+&{C_{B.12}\over s_{34}s_{234}l^2(l-p_1)^2}+{C_{C.1}\over s_{12}(l+p_2)^2(l+p_2+p_3)^2(l-p_1)^2}+{C_{C.5}\over s_{14}(l+p_4)^2(l+p_2+p_4)^2(l-p_1)^2}\nn
 &+&{C_{C.8}\over s_{23}l^2(l+p_2+p_3)^2(l-p_1)^2}+{C_{C.12}\over s_{34}l^2(l+p_2)^2(l-p_1)^2}+{C_{D.1}\over l^2(l+p_3)^2(l+p_3+p_4)^2(l-p_1)^2},
 \eea
 where we have neglect  terms integrated to zero. The above expression is nothing but four-point one-loop integrand $\W I(1,2,3,4)$ of color-ordered scalar theory . After similar calculations for other ordering, we do get the claimed form \eref{1loop-Dual-DDM}
 \bea
 I(1,2,3,4)(l)=\Sl_{\sigma\in permutations~of \{2,3,4\}}n_{1,\sigma}(l)\W I(1,\sigma).
 \eea
up to terms vanishing after loop integration, where each $\W I(1,\sigma)$ is identified to the integrand of color-ordered scalar theory at one-loop. For higher points, the general procedure is same although computations will be much more complicated.

%%%%%%%%%%%%%%%%%%%%%%%%%%%%%%%%
\section{Dual trace-form}
%%%%%%%%%%%%%%%%%%%%%%%%%%%%%%%%
\label{sec:dual-trace}
In the discussions above we saw that the dual DDM-form can be derived
through relatively straightforward manipulations. Deriving
a corresponding dual trace-form at one-loop however turns out to be less direct,
especially
because of the extra conditions required to define dual trace factors \cite{Bern:2011ia, Du:2013sha}.

Recall that at tree-level, the set of numerators $n_{1\sigma n}$, consisting of $(n-2)!$ elements,
having legs $1$ and $n$ fixed at two ends, were translated into $(n-1)!$ dual traces $\tau_{1\W\sigma}$,
which are counterparts of the single color trace factors. To uniquely determine  $\tau$ we need to
impose  KK-relation among $\tau_{1\W\sigma}$'s, so the number of independent dual traces can be reduced to $(n-2)!$. The algorithm formally picks a fixed pair $(1,n)$ to define
basis numerators. To examine if the solution satisfy relabling symmetry
we need to inspect the transformation under permutaions of legs $1$ and $n$.

At one-loop level, similar constraints are required to properly define dual traces.
For the purpose of discussion let us first review the  $U(N_{c})$ color structure at one-loop,
which also serves as input to the definition of dual traces.

%%%%%%%%%%%%%%%%%%%%%%%%%%%%%%%%%%%
\subsection{general structure of the defining conditions}
%%%%inspiration from U(N) color algebra\\

Generically, the color factors appear in the DDM-form
at one-loop level can be translated into double trace factors,
\bea
c^{1-loop}_{\{\sigma\}}&=&f^{x_1a_1x_2}f^{x_2a_2x_3}\dots f^{x^na_nx_1}=\Tr(T^{x_1}[T^{a_1},[T^{a_2},\dots[T^a_n,T^{x_1}]]])\nn
&=&\Sl_{\sigma\in OP(\{\alpha\}\bigcup\{\beta\})}(-1)^{n_{\beta}}\Tr(T^{x_1}T^{a_{\alpha_1}}\dots T^{a_{\alpha_{n_{\alpha}}}}T^{x_1}T^{\beta_{n_{\beta}}}\dots T^{\beta_1})\nn
&=&\Sl_{\sigma\in OP(\{\alpha\}\bigcup\{\beta\})}(-1)^{n_{\beta}}\Tr(T^{a_{\alpha_1}}\dots T^{a_{\alpha_{n_{\alpha}}}})\Tr(T^{\beta_{n_{\beta}}}\dots T^{\beta_1}),~~\Label{c-Tr}
\eea
%
%%At the third line, the general term will be
where in the last line we used the property of $U(N_{c})$,
\bea \sum_{x_1} {\rm Tr}( T^{x_1}AT^{x_1} B)={\rm Tr} (A) {\rm Tr} (B). \eea
Note however, two exceptional cases call for special attention.
When the repeated generators are adjacent, single trace factors are produced instead.
This can happen in equation (\ref{c-Tr}) as
%% i.e., we will get the double trace. However, there are two special case. The first one is
%
\bea \sum_{x_1}{\rm Tr} ( T^{x_1} T^{a_1}.... T^{a_n} T^{x_1})=  N_c{\rm Tr} (  T^{a_1}.... T^{a_n} )\eea
%
%%%and the second term is
or as
\bea (-)^n\sum_{x_1}{\rm Tr} ( T^{x_1} T^{x_1}T^{a_n} T^{a_{n-1}}.... T^{a_1} )=  (-)^n
N_c{\rm Tr} (  T^{a_n} T^{a_{n-1}}.... T^{a_1} ).\eea

Inspired by the above algebraic structure, it is natural to assume that there
are kinematic correspondence of the following color trace factors
\bea   {\rm Tr}( T^{\a_1}...T^{\a_{m}}) {\rm Tr}( T^{\b_1}... T^{\b_{n}}) \to \tau_{\a;\b},~~~
{\rm Tr} (  T^{a_1}.... T^{a_n} )
\to \tau_{ \a }, ~~~\Label{map-1}\eea
where kinematic trace factors $\tau_{\a;\b}$ and $\tau_{ \a }$ are cyclic invariant.
Thus we can impose following relation between $n_\a$ in dual DDM-form and kinematic trace structure
$\tau_{\a;\b}$ in dual trace-form as
\bea
n^{1-loop}_{\{\sigma\}}
&=&\Sl_{\sigma\in OP(\{\alpha\}\bigcup\{\beta\})}(-1)^{n_{\beta}}\tau_{\a;\b^T},~~\Label{n-tau-rel}
\eea
where $\b^T$ means reversing the ordering in the subset $\b$. In \eref{n-tau-rel}, there are again
two special cases: when $\a=\emptyset$, $\tau_{\a;\b^T}\to N_c \tau_{\b^T}$ and when $\b=\emptyset$, $\tau_{\a;\b^T}\to N_c \tau_{\a}$. In other words, there are two single traces and we have also kept the
possible freedom of "kinematic rank $N_c$".

From equation \eref{n-tau-rel}, one can see that the number of all $\tau$ together
clearly exceeds that of $n$. In fact, there are only $(n-1)!/2$ independent $n_\a$'s because
$n_\a$ is cyclic invariant and satisfies following reflection relation
\bea n_{\a}=(-)^{M} n_{\a^T}~~~~~\Label{n-reverse}\eea
where $M$ is the number  of elements of the  set $\a$. To be able to solve $\tau$ by $n_\a$, we need to impose
extra equations.  In viewing of the solution that works at tree-level, a natural generalization is to impose the one-loop KK-relation \cite{Bern:2011ia} between kinematic
 single and double trace parts
\bea \tau_{\a;\b}=(-)^{n_\b}\sum_{C\in Z_{n_\b}}\sum_{\sigma\in {\cal COP}( \a \bigcup C(\b)^T)}\tau_{\sigma}
~~~~\Label{map-2}\eea
It is worth to notice that in \eref{map-2}, subsets $\a$ and $\b$ have been treated differently at the right handed side. However, for double trace part, by the correspondence to color part it is very naturally to impose
\bea \tau_{\a;\b}=\tau_{\b;\a}~~~~\Label{tau-double-sym}\eea
Thus, to be consistence between \eref{map-2} and \eref{tau-double-sym}, we need to impose reflection relation
 in addition
\bea
\tau_{\a}=(-1)^M\tau_{\a^T}.~~~~~~~~\Label{tau-reverse}
\eea
where $M$ is the number of elements of the set $\a$. With these extra conditions \eref{map-2}, \eref{tau-double-sym} and \eref{tau-reverse}, the number of independent kinematic trace factors
is reduced $(n-1)!/2$. So finally  the original equations (\ref{n-tau-rel})
become an $(n-1)!/2$ by $(n-1)!/2$ matrix equation,
\bea n_{1\sigma}=\sum_{\sigma'\in S_{n-1}} G[\sigma|\sigma'] \tau_{1\sigma'}~~~~\Label{n-tau-final}\eea
Knowing the matrix $G$, we can solve $\tau_{1\sigma'}$ by $n_{1\sigma}$ and finally determine all
kinematic trace factors.

%%%%%%%%%%%%%%%%%%%%%%%%
\subsection*{general algorithm}
%%%%%%%%%%%%%%%%%%%%%%%%%%
To summarize, the general algorithm of constructing kinematic trace factors  is given by the following:
\begin{itemize}
%%%
%%\item
\item Starting with any dual-DDM basis numerator $n_{1,...,n}$ we consider all possible splittings of its label $\{1,...,n\}$ into two subsets $\a, \b$, each can be empty.  Generically there will be $2^n$ splittings.
For example at four-points, denoting the one-loop dual-DDM factor as $n^{1-loop}_{1,\sigma}$, $\sigma\in perm(2,3,4)$. the relation between $n$ and $\tau$ is given by
\bea
n^{1-loop}_{1234}&=& N_{c} \tau_{\{1234\}}-\tau_{\{234\},\{1\}}-\tau_{\{134\},\{2\}}-\tau_{\{124\},\{3\}}-\tau_{\{123\},\{4\}}\nn
&&+\tau_{\{34\},\{21\}}+\tau_{\{24\},\{31\}}+\tau_{\{23\},\{41\}}+\tau_{\{13\},\{42\}}+\tau_{\{14\},\{32\}}
+\tau_{\{12\},\{43\}}\nn
&&-\tau_{\{4\},\{321\}}-\tau_{\{3\},\{421\}}-\tau_{\{2\},\{431\}}-\tau_{\{1\},\{432\}}+
N_{c} \tau_{\{4321\}}\Label{4-pt}
\eea
%

%\item Because cyclic symmetry, we can set one element, for example, leg $1$,
% to be at the first position in set $\a$.

%
\item We then impose KK relation on $\tau$ \eref{map-2}. In the four-point case, we have
\bea
\tau_{\{bcd\},\{a\}}&=&-\tau_{\{abcd\}}-\tau_{\{bacd\}}-\tau_{\{bcad\}},\nn
\tau_{\{cd\},\{ba\}}&=&\tau_{\{abcd\}}+\tau_{\{acbd\}}+\tau_{\{cabd\}}+\tau_{\{bacd\}}+\tau_{\{bcad\}}
+\tau_{\{cbad\}},\nn
\tau_{\{d\},\{cba\}}&=&-\tau_{\{abcd\}}-\tau_{\{abdc\}}-\tau_{\{adbc\}},
\eea
Substituting these relations into \eqref{4-pt} and using cyclic symmetry  $\tau_{\{abcd\}}=\tau_{\{dabc\}}$, we get
\bea
n^{1-loop}_{1234}=(15+N_{c})\tau_{\{1234\}}+10\tau_{\{1243\}}+10\tau_{\{1324\}}
+10\tau_{\{1342\}}+10\tau_{\{1423\}}+(5+N_{c})\tau_{\{1432\}}.~~~\Label{n-tau}
\eea
\item Using reflection relation \eref{tau-reverse} we can reduce the obtained equations further. For
example, above equation is reduced to
\bea n^{1-loop}_{1234}&=&(20+2 N_{c})\tau_{1234}+20\tau_{1243}+20\tau_{1324},~~~\Label{4-point-rel}\eea
Repeating the same manipulations for all basis numerators, we arrive at the matrix equation (\ref{n-tau-final}), from which we can solve for all dual traces.

\end{itemize}
%

%%%%%%%%%%%%%%%%%%%%%%%%%%%
%%%%%%%%%%%%%%%%%%%%%%%%
\subsection*{$G$-matrix:}
%%%%%%%%%%%%%%%%%%%%%%%%%%

Now we discuss the computation of $G$-matrix. The calculation can be divided into two steps.
 The first step is to calculate extended $\W G$-matrix $\W G[\sigma|\rho]$ where $\sigma,\rho\in S_n/Z_n$ (i.e.,
 all permutations up to cyclic ordering).
 The second step is to impose the reflection relation \eref{tau-reverse}, i.e,
 \bea  G[\sigma|\rho]=\W G[\sigma|\rho]+(-)^n \W G[\sigma|\rho^T],~~~~~\sigma,\rho\in (S_n/Z_n)/Z_2~~~~\Label{G-gen}\eea
Since the second step is easy, we will focus on the first step, i.e., the extended $\W G$-matrix.
Elements of extended $\W G$-matrix depend on $N_c$ only for following two kinds of structures
\bea \W G[\sigma|\sigma]= a_0+ N_c,~~~~~~~\W G[\sigma|\sigma^T]=(-)^n( b_0+ N_c)\eea
where $a_0,b_0$ are constants. Because this dependence, if we know the  extended $\W G$-matrix for $N_c=1$,
we will know the  extended $\W G$-matrix for general $N_c$.

To demonstrate the calculation of element $\W G[\sigma|\rho]$, let us use four-point result \eref{n-tau}
with $N_c=1$ as an
example. For this example, we have $\sigma=\{1,2,3,4\}$ fixed and $3!=6$ different choices of $\rho$. Given
the ordering of $\sigma$, there are $2^4=16$ different splittings to two subsets. Among them, $8$ of them with $1$ at the first subset are given by (remembering to keep relative ordering)
\bea & & \{1,2,3,4\}\to (\sigma_L, \sigma_R)\nn &=  & (1234,\emptyset)/(123,4)/(124,3)/(134,2)/(12,34)/(13,24)/(14,23)/(1,234)~~~\Label{sigma-split}\eea
and other $8$ are obtained by exchanging these two subsets. We do similar splitting  to the ordering $\rho$,
but now we will allow the cyclic shifting of one subset. For example, with $\rho=\{1,2,4,3\}$ we will have
following splitting with $1$ at the first position of the first subset (by cyclic symmetry, we can always
fix one element)
\bea \{1,2,4,3\}\to (\rho_L,\rho_R) & = & (1243, \emptyset)/(124,3)/(123,4)/(143,2)/(12,43)/(12,34)/(14,23)/(14,32)\nn & &
/(13,24)/(13,42)/(1,243)/(1,432)/(1,324)~~~\Label{rho-split} \eea
where since we have fixed $1$, we have to include the cyclic shifting of $\rho_R$. Comparing these two
splitting \eref{sigma-split} and \eref{rho-split}, we see that there are five splittings to be same:
\bea (123,4)/(124,3)/(12,34)/(13,24)/(14,23)  \Longrightarrow \W G[\{1,2,3,4\}|\{1,2,4,3\}]
=2\times 5=10\eea
where factor $2$ comes from exchanging of two subsets. One can easily check that all other five coefficients
in \eref{n-tau} can be obtained by same way. For $\rho=\{1,2,3,4\}$ there are $8\times 2$ splitting. For
$\rho= \{1,3,2,4\}$,  $(124,3)/(134,2)/(12,34)/(13,24)/(14,23)$ from \eref{sigma-split} are taken, so $5\times 2=10$. For
$\rho= \{1,3,4,2\}$,  $(134,2)/(12,34)/(13,24)/(14,23)/(1,234)$ from \eref{sigma-split} are taken, so $5\times 2=10$. For
$\rho= \{1,4,2,3\}$,  $(123,4)/(12,34)/(13,24)/(14,23)/(1,234)$ from \eref{sigma-split} are taken, so $5\times 2=10$. Finally for
$\rho= \{1,4,3,2\}$,  $(12,34)/(13,24)/(14,23)$ from \eref{sigma-split} are taken, so $3\times 2=6$.

Having about general discussions, now we demonstrate our algorithm by several examples.

%%%%%%%%%%%%%%%%%%%%%%
%%%%%%%%%%%%%%%%%%%%%%%
%%%
%%% 4-pt example\\
%%%
%%% general algorithm\\

%%%%%%%%%%%%%%%%%%%%%%
%%%%%%%%%%%%%%%%%%%%%%%
%%%%%%%%%%%%%%%%%%%%%%%%%

%%%%%%%%%%%%%%%%%%%%
\subsection{Four-point dual traces}
%%%%%%%%%%%%%%%%%%%%

Under our imposed conditions \eref{map-2}, \eref{tau-double-sym} and \eref{tau-reverse}
 the number of independent $n$'s and $\tau$'s is  ${(4-1)!\over 2}=3$. We take the liberty to choose following
 three orderings  $(1234)$, $(1243)$ and $(1324)$ as our basis. Using our algorithm for $G$-matrix,
equation (\ref{n-tau-final})  yields
\bea
n^{1-loop}_{1234}&=&(20+2 N_{c})\tau_{1234}+20\tau_{1243}+20\tau_{1324},\nn
n^{1-loop}_{1243}&=&20\tau_{1234}+(20+2 N_{c})\tau_{1243}+20\tau_{1324},\nn
n^{1-loop}_{1324}&=&20\tau_{1234}+22\tau_{1243}+(20+2 N_{c})\tau_{1324}.
\eea
The  determinant of $G$-matrix is  ${\rm det}(G)=8 N_c^2 (30+N_c)$ for generic $N_{c}$,
from which we derive the solution for $\tau_{1234}$,
\bea
\tau_{1234}=\frac{1}{2N_{c} (30+N_{c})} \left( (20+N_{c}) n_{1234}- 10 n_{1243}- 10 n_{1324}
\right).~~~\Label{4pt-tau}
\eea
Expressions of other orderings  $\tau_\rho$ can be obtained by relabeling symmetry.

This result seems to differ  from the  result previously obtained for $N=4$ SYM theory in \cite{Bern:2011ia}.
To connect the two results, notice that for $N=4$ SYM, only have box diagrams contribute and the corresponding $n$ is given as
\bea
n^{1-loop}_{abcd}=s_{ab}s_{ad}A^{tree}(a,b,c,d).~~\Label{4pt-N=4-n}
\eea
Substituting \eqref{4pt-N=4-n} into \eqref{4pt-tau} and using tree-level amplitude relation to write all the four-point tree amplitudes in terms of $A(1234)$, we get
\bea
\tau^{1-loop}_{1234}={1\over 62}stA^{tree}(1,2,3,4).
\eea
which is just the result given by Bern and Dennen when $N_{c}$ is chosen to be $1$.

Our result \eref{4pt-tau} has a free parameter $N_c$. It is easy to see that $N_c$ can not be $0$ or $-30$
because for these two values, determinant of $G$-matrix is zero, i.e., $G$-matrix is degenerated and we can not
solve $\tau$ by $n_\a$. Also, with particular choice of $N_c$, we may get simpler expressions. For example,
when $N_c=-10$ we get
\bea \tau_{1234}=\frac{-1}{40} \left( + n_{1234}-  n_{1243}- n_{1324}
\right)\eea
while when $N_c=-20$ we get
\bea \tau_{1234}=\frac{1}{40} \left(  n_{1243}+ n_{1324}
\right)\eea
%

%%%%%%%%%%%%%%%%%%%%%%%%%%%%%%%
\subsection{Five-point case}
%%%%%%%%%%%%%%%%%%%%%%%%%%%%%%%
\label{sec:5-pt-example}
Let us apply the same algorithm to five points. The relation between DDM basis numerator $n$ and dual trace $\tau$
is given as
\bea n_{a_1 a_2 a_3 a_4 a_5} & = & (30+2 N_c) \tau_{\{ a_1 a_2 a_3 a_4 a_5 \} }
+12 \tau_{\{ a_1 a_2 a_3 a_5 a_4 \} }+ 12 \tau_{\{ a_1 a_2 a_4 a_3 a_5 \} }
+12\tau_{\{ a_1 a_3 a_2 a_4 a_5 \} }
\nn
& & + 6 \tau_{\{ a_1 a_2 a_4 a_5 a_3 \} }+ 6 \tau_{\{ a_1 a_2 a_5 a_3 a_4 \} }
+6 \tau_{\{ a_1 a_3 a_4 a_2 a_5 \} }+ 6 \tau_{\{ a_1 a_4 a_2 a_3 a_5 \} }
\nn
& & -12 \tau_{\{ a_1 a_2 a_5 a_4 a_3 \} }-12 \tau_{\{ a_1 a_4 a_3 a_2 a_5 \} }
 -6 \tau_{\{ a_1 a_3 a_2 a_5 a_4 \} }+ 0 \tau_{\{ a_1 a_3 a_5 a_2 a_4 \} }~~~~\Label{5-point-rel}\eea
The number of independent $n$s and $\tau$s is
 $(5-1)!/ 2=12$. We choose the our basis to be
 $(12345)$, $(12354)$, $(12435)$, $(12453)$, $(12434)$, $(12543)$, $(13245)$, $(13254)$, $(13425)$, $(13524)$, $(14235)$, $(14325)$ in order,
 which leads to following  matrix $G$
%
%%%{\tiny
\bea
{\tiny
G=\left(
\begin{array}{cccccccccccc}
 2 \text{Nc}+30 & 12 & 12 & 6 & 6 & -12 & 12 & -6 & 6 & 0 & 6 & -12 \\
 12 & 2 \text{Nc}+30 & 6 & -12 & 12 & 6 & -6 & 12 & 0 & 6 & 12 & -6 \\
 12 & 6 & 2 \text{Nc}+30 & 12 & -12 & 6 & 6 & 0 & -12 & 6 & 12 & 6 \\
 6 & -12 & 12 & 2 \text{Nc}+30 & 6 & 12 & 12 & -6 & -6 & -12 & -6 & 0 \\
 6 & 12 & -12 & 6 & 2 \text{Nc}+30 & 12 & 0 & 6 & 6 & -12 & -6 & -12 \\
 -12 & 6 & 6 & 12 & 12 & 2 \text{Nc}+30 & -6 & 12 & -12 & -6 & 0 & 6 \\
 12 & -6 & 6 & 12 & 0 & -6 & 2 \text{Nc}+30 & 12 & 12 & 6 & -12 & 6 \\
 -6 & 12 & 0 & -6 & 6 & 12 & 12 & 2 \text{Nc}+30 & 6 & 12 & -6 & 12 \\
 6 & 0 & -12 & -6 & 6 & -12 & 12 & 6 & 2 \text{Nc}+30 & -12 & 6 & 12 \\
 0 & 6 & 6 & -12 & -12 & -6 & 6 & 12 & -12 & 2 \text{Nc}+30 & -12 & -6 \\
 6 & 12 & 12 & -6 & -6 & 0 & -12 & -6 & 6 & -12 & 2 \text{Nc}+30 & 12 \\
 -12 & -6 & 6 & 0 & -12 & 6 & 6 & 12 & 12 & -6 & 12 & 2 \text{Nc}+30
\end{array}
\right) } \eea %%%}
with determinant ${\rm det}(G)=2^{12} N_c^6 (N_c+30)^6$. Therefore the solution is
\bea & & \tau_{\{1 2 3 4 5\}}  =
{1\over 2 N_c (30+N_c)}
\left\{ (15+ N_c) n_{1 2 3 4 5} -6
n_{1 2 3 5 4} -6 n_{1 2 4 3 5}
-3 n_{1 2 4 5 3}\right. \nn & & \left.
-3 n_{1 2 5 3 4}+ 6
n_{1 2 5 4 3} -6 n_{1 3 2 4 5}
+3  n_{1 3 2 5 4}
-3 n_{1 3 4 2 5}-3
n_{1 4 2 3 5} +6 n_{1 4 3 2 5} \right\}\eea
Other $\tau$'s can be obtained using relabeling symmetry. For this expression, if we choose $N_c=-15$,
all coefficients are ${\pm 2\over 150}$ and ${1\over 150}$. Especially
the first coefficients $(15+N_c)\to 0$.

%%%%%%%%%%%%%%%%%%%%%%%
\subsection{Six-point example}
%%%%%%%%%%%%%%%%%%%%%%%%%%%
At six-points,  the basis can be labeled by
the following $(6-1)!/2=60$ orderings
\bea
& & \{1, 2, 3, 4, 5, 6\}, \{1, 2, 3, 4, 6, 5\}, \{1, 2, 3, 5, 4, 6\}, \{1, 2, 3,
   5, 6, 4\}, \{1, 2, 3, 6, 4, 5\}, \{1, 2, 3, 6, 5, 4\},\nn
   & &  \{1, 2, 4, 3, 5,
  6\}, \{1, 2, 4, 3, 6, 5\}, \{1, 2, 4, 5, 3, 6\}, \{1, 2, 4, 5, 6, 3\}, \{1,
  2, 4, 6, 3, 5\}, \{1, 2, 4, 6, 5, 3\},\nn
  & &  \{1, 2, 5, 3, 4, 6\}, \{1, 2, 5, 3,
   6, 4\}, \{1, 2, 5, 4, 3, 6\}, \{1, 2, 5, 4, 6, 3\}, \{1, 2, 5, 6, 3,
  4\}, \{1, 2, 5, 6, 4, 3\},\nn
  & &  \{1, 2, 6, 3, 4, 5\}, \{1, 2, 6, 3, 5, 4\}, \{1,
  2, 6, 4, 3, 5\}, \{1, 2, 6, 4, 5, 3\}, \{1, 2, 6, 5, 3, 4\}, \{1, 2, 6, 5,
   4, 3\}, \nn
   & & \{1, 3, 2, 4, 5, 6\}, \{1, 3, 2, 4, 6, 5\}, \{1, 3, 2, 5, 4,
  6\}, \{1, 3, 2, 5, 6, 4\}, \{1, 3, 2, 6, 4, 5\}, \{1, 3, 2, 6, 5, 4\},\nn
  & &  \{1,
  3, 4, 2, 5, 6\}, \{1, 3, 4, 2, 6, 5\}, \{1, 3, 4, 5, 2, 6\}, \{1, 3, 4, 6,
   2, 5\}, \{1, 3, 5, 2, 4, 6\}, \{1, 3, 5, 2, 6, 4\},\nn
   & &  \{1, 3, 5, 4, 2,
  6\}, \{1, 3, 5, 6, 2, 4\}, \{1, 3, 6, 2, 4, 5\}, \{1, 3, 6, 2, 5, 4\}, \{1,
  3, 6, 4, 2, 5\}, \{1, 3, 6, 5, 2, 4\}, \nn
  & & \{1, 4, 2, 3, 5, 6\}, \{1, 4, 2, 3,
   6, 5\}, \{1, 4, 2, 5, 3, 6\}, \{1, 4, 2, 6, 3, 5\}, \{1, 4, 3, 2, 5,
  6\}, \{1, 4, 3, 2, 6, 5\},\nn
  & &  \{1, 4, 3, 5, 2, 6\}, \{1, 4, 3, 6, 2, 5\}, \{1,
  4, 5, 2, 3, 6\}, \{1, 4, 5, 3, 2, 6\}, \{1, 4, 6, 2, 3, 5\}, \{1, 4, 6, 3,
   2, 5\}, \nn
   & & \{1, 5, 2, 3, 4, 6\}, \{1, 5, 2, 4, 3, 6\}, \{1, 5, 3, 2, 4,
  6\}, \{1, 5, 3, 4, 2, 6\}, \{1, 5, 4, 2, 3, 6\}, \{1, 5, 4, 3, 2, 6\}~~~~~~~~\eea
The expansion coefficients of $n_{123456}$ into $\tau$s, i.e, $G[\{123456\}|\rho]=G_{1i}$ $i=1,...,60$ respectively by the orderings listed above, is given by
\bea  G_{1i}=\{62 + 2   N_c, & &  34, 34, 22, 22, 12, 34, 22, 22, 22,\nn
  18, & & 16, 22, 18, 12,
16, 26, 22, 22, 16, \nn
 16,& & 22, 22, 34, 34, 22, 22, 16, 16, 22, \nn
 22,& & 16,
22, 18, 18, 12, 16, 18, 18, 18, \nn
 12,& & 18, 22, 16, 18, 12, 12, 22, 16,
18, \nn
26,& & 22, 18, 18, 22, 16, 16, 22, 22, 34\},\eea
Other $G_{ij}$ can be obtained by relabeling symmetry. The determinant of  matrix $G$ is
\bea {\rm det}(G)=2^{60} N_c^{24}(N_c+18)^{5} (N_c+21)^{16}
(N_c+56)^9 (N_c+60)^5 (N_c+630),\eea
and the solution is given by
\bea \tau_{\sigma}= G^{-1}[\sigma|\rho] n_{\rho}\eea
The inverse of matrix $G$ is very complicated, but with relabeling symmetry, it is enough to give the
first row, i.e., $G^{-1}_{1i}$ with $i=1,...,60$. To have a feeling about the $N_c$-dependence, we list
all $60$ elements as following:
\bea G^{-1}_{11,12,13,14,15} &= & \Biggl\{ \frac{1}{120}
   \left(\frac{5}{N_c+18}+\frac{16}{N_c+21}+\frac{9}{N_c+56}+\frac{
   5}{N_c+60}+\frac{1}{N_c+630}+\frac{24}{N_c}\right), \nn
   & & {1\over 2520} \left(-\frac{35}{N_c+18}+\frac{56}{N_c+21}+\frac{99}{N_c+56}+\frac{
   35}{N_c+60}+\frac{21}{N_c+630}-\frac{176}{N_c}\right), \nn
   & & {1\over 2520} \left({-\frac{35}{N_c+18}+\frac{56}{N_c+21}+\frac{99}{N_c+56}+\frac{35}{\text{N
   c}+60}+\frac{21}{N_c+630}-\frac{176}{N_c}}\right),\nn
   & & {1\over 2520} \left({\frac{35}{N_c+18}+\frac{28}
   {N_c+21}+\frac{27}{N_c+56}-\frac{35}{N_c+60}+\frac{21}{N_c+630}-\frac{76}
   {N_c}}\right),\nn
   & & {1\over 2520}\left({\frac{35}{N_c+18}+\frac{28}{N_c+21}+\frac{27}{N_c+56}-\frac{
   35}{N_c+60}+\frac{21}{N_c+630}-\frac{76}{N_c}}\right)\Biggr\}\eea
%
%%%%%%%%%%%%%%%%%%%%%%%%%%%%%%%%%
%
\bea G^{-1}_{16,17,18,19,1(10)} &= &\Biggl\{{1\over 2520}\left(-\frac{105}{N_c+18}+\frac{45}{N_c+56}-\frac{105}{N_c+60}+\frac{21}{
   N_c+630}+\frac{144}{N_c}\right),\nn
 & &{1\over 2520}\left(-\frac{35}{N_c+18}+\frac{56}{N_c+21}+\frac{99}{N_c+56}
  +\frac{35}{N_c+60}+\frac{21}{N_c+630}-\frac{176}{N_c}  \right),\nn
   & & {1\over 2520}\left({-\frac{35}{N_c
   +18}-\frac{112}{N_c+21}+\frac{27}{N_c+56}+\frac{35}{N_c+60}+\frac{21}{N_c+63
   0}+\frac{64}{N_c}}\right),\nn
   & & {1\over 2520}\left({\frac{35}{N_c+18}+\frac{28}{N_c+21}+\frac{27}{\text
   {Nc}+56}-\frac{35}{N_c+60}+\frac{21}{N_c+630}-\frac{76}{N_c}}\right),\nn
   & & {1\over 2520}\left({\frac{35
   }{N_c+18}+\frac{28}{N_c+21}+\frac{27}{N_c+56}-\frac{35}{N_c+60}
   +\frac{21}{N_c+630}-\frac{76}{N_c}}\right)\Biggr\}\eea
%
%%%%%%%%%%%%%%%%%%%%%%
%
\bea G^{-1}_{1(11),1(12),1(13),1(14),1(15)} &= &\Biggl\{{1\over 2520}\left({-\frac{35}{N_c+18}+\frac{56}{N_c+21}-\frac{81}{N_c+56}+\frac{35}{\text{
   Nc}+60}+\frac{21}{N_c+630}+\frac{4}{N_c}}\right),\nn
   & &{1\over 2520}\left(\frac{35}{N_c+18}-\frac{56}{N_c+21}-\frac{9}{N_c+56}-\frac{35}{N_c+60}+\frac{2
   1}{N_c+630}+\frac{44}{N_c}\right),\nn
   & & {1\over 2520}\left({\frac{35}{N_c+18}+\frac{28}{N_c+21}+\frac{27}{N_c+56}-\frac{35}
   {N_c+60}+\frac{21}{N_c+630}-\frac{76}{N_c}}\right),\nn
   & & {1\over 2520}\left({-\frac{35}{N_c
   +18}+\frac{56}{N_c+21}-\frac{81}{N_c+56}+\frac{35}{N_c+60}+\frac{21}{N_c+63
   0}+\frac{4}{N_c}}\right),\nn
   & &{1\over 2520}\left( -\frac{105}{N_c+18}+\frac{45}{N_c+56}-\frac{105}{N_c+60}+\frac{21}{N_c+630}
   +\frac{144}{N_c}\right)\Biggr\} \eea
%
%%%%%%%%%%%%%%%%%%%%%%%%%%%
%
\bea G^{-1}_{1(16),1(17),1(18),1(19),1(20)} &= & \Biggl\{{1\over 2520}\left(\frac{35}{N_c+18}-\frac{56}{N_c+21}-\frac{9}{N_c+56}-\frac{35}{N_c+60}
+\frac{21}{N_c+630}+\frac{44}{N_c}\right),\nn
   & & \frac{1}{840}
   \left(\frac{35}{N_c+18}-\frac{56}{N_c+21}-\frac{9}{N_c+56}+\frac{35}{N_c+60
   }+\frac{7}{N_c+630}-\frac{12}{N_c}\right),\nn
   & & {1\over 2520}\left({-\frac{35}{N_c+18}-\frac{112}{N_c+21}+\frac{27}{N_c+56}
   +\frac{35}{N_c+60}+\frac{21}{N_c+630}+\frac{64}{N_c}}\right),\nn
   & & {1\over 2520}\left({\frac{35}{N_c+18}+\frac{28}{N_c+21}+\frac{27}{N_c+56}-\frac{35
   }{N_c+60}+\frac{21}{N_c+630}-\frac{76}{N_c}}\right),\nn
   & & {1\over 2520}\left(\frac{35}{N_c+18}-\frac{56}{N_c+21}-\frac{9}{N_c+56}-\frac{35}{N_c+60}+\frac{2
   1}{N_c+630}+\frac{44}{N_c}\right)\Biggr\} ~~~~~~ \eea
%
%%%%%%%%%%%%%%%%%%%%%%%%%%%
%
\bea G^{-1}_{1(21),1(22),1(23),1(24),1(25)} &= &\Biggl\{{1\over 2520}\left(\frac{35}{N_c+18}-\frac{56}{N_c+21}-\frac{9}{N_c+56}-\frac{35}{N_c+60}+\frac{21
   }{N_c+630}+\frac{44}{N_c}\right),\nn
   & &{1\over 2520}\left({-\frac{35}{N_c+18}-\frac{112}{N_c+21}+\frac{27}{N_c+56}+
   \frac{35}{N_c+60}+\frac{21}{N_c+630}+\frac{64}{N_c}}\right),\nn
   & & {1\over 2520}\left({-\frac{35}{N_c
   +18}-\frac{112}{N_c+21}+\frac{27}{N_c+56}+\frac{35}{N_c+60}+\frac{21}{N_c+63
   0}+\frac{64}{N_c}}\right),\nn
   & & {1\over 2520}\left({-\frac{35}{N_c+18}+\frac{56}{N_c+21}+\frac{99}{N_c+56}
   +\frac{35}{N_c+60}+\frac{21}{N_c+630}-\frac{176}{N_c}}\right),\nn
   & & {1\over 2520}\left({-\frac
   {35}{N_c+18}+\frac{56}{N_c+21}+\frac{99}{N_c+56}+\frac{35}{N_c+60}+\frac{21}
   {N_c+630}-\frac{176}{N_c}}\right)\Biggr\}~~~~~~~ \eea

%%%%%%%%%%%%%%%%%%%%%%%%%%%
%
\bea G^{-1}_{1(26),1(27),1(28),1(29),1(30)} &= & \Biggl\{{1\over 2520}\left({-\frac{35}{N_c+18}-\frac{112}{N_c+21}+\frac{27}{N_c+56}+\frac{35}{\text{
   Nc}+60}+\frac{21}{N_c+630}+\frac{64}{N_c}}\right),\nn
   & & {1\over 2520}\left({-\frac{35}{N_c+18}-\frac{11
   2}{N_c+21}+\frac{27}{N_c+56}+\frac{35}{N_c+60}+\frac{21}{N_c+630}+\frac{64}{
   N_c}}\right),\nn
   & & {1\over 2520}\left(\frac{35}{N_c+18}-\frac{56}{N_c+21}-\frac{9}{N_c+56}-\frac{35}{N_c+60}+\frac{21
   }{N_c+630}+\frac{44}{N_c}\right),\nn
   & &{1\over 2520}\left(\frac{35}{N_c+18}-\frac{56}{N_c+21}-\frac{9}{N_c+56}-\frac{35}{N_c+60}+\frac{21
   }{N_c+630}+\frac{44}{N_c}\right),\nn
   & & {1\over 2520}\left({\frac{35}{N_c+18}+\frac{28}{N_c+21}+\frac{27}{N_c+56}-\frac
   {35}{N_c+60}+\frac{21}{N_c+630}-\frac{76}{N_c}}\right)\Biggr\}~~~~~~~\eea
%
%%%%%%%%%%%%%%%%%%%%%%%%%%%
%
\bea G^{-1}_{1(31),1(32),1(33),1(34),1(35)} &= &
\Biggl\{{1\over 2520}\left({\frac{35}{N_c+18}+\frac{28}{N_c+21}+\frac{27}{N_c+56}-\frac{35}{\text{Nc
  }+60}+\frac{21}{N_c+630}-\frac{76}{N_c}}\right),\nn
& & {1\over 2520}\left(\frac{35}{N_c+18}-\frac{56}{N_c+21}-\frac{9}{N_c+56}-\frac{35}{N_c+60}
  +\frac{21
  }{N_c+630}+\frac{44}{N_c}\right),\nn
  & & {1\over 2520}\left({\frac{35}{N_c+18}+\frac{28}{N_c+21}+\frac{27}{N_c+56}-
 \frac{35}{N_c+60}+\frac{21}{N_c+630}-\frac{76}{N_c}}\right),\nn
   & & {1\over 2520}\left({-\frac{35}{N_c+1
   8}+\frac{56}{N_c+21}-\frac{81}{N_c+56}+\frac{35}{N_c+60}+\frac{21}{N_c+630}+
   \frac{4}{N_c}}\right),\nn
   & & {1\over 2520}\left({-\frac{35}{N_c+18}+\frac{56}{N_c+21}-\frac{81}{\text{Nc
   }+56}+\frac{35}{N_c+60}+\frac{21}{N_c+630}+\frac{4}{N_c}}\right)\Biggr\}~~~~~~\eea
%

%%%%%%%%%%%%%%%%%%%%%%%%%%%
%
\bea G^{-1}_{1(36),1(37),1(38),1(39),1(40)} &= & \Biggl\{{1\over 2520}\left(\frac{35}{N_c+18}+\frac{112}{N_c+21}-\frac{117}{N_c+56}-\frac{35}{N_c+60}+\frac
   {21}{N_c+630}-\frac{16}{N_c}\right),\nn
   & &{1\over 2520}\left( \frac{35}{N_c+18}-\frac{56}{N_c+21}-\frac{9}{N_c+56}-\frac{35}{N_c+60}+\frac{21
   }{N_c+630}+\frac{44}{N_c}\right),\nn
   & &{1\over 2520}\left({-\frac{35}{N_c+18}+\frac{56}{N_c+21}-\frac{81}{N_c+56}+
   \frac{35}{N_c+60}+\frac{21}{N_c+630}+\frac{4}{N_c}}\right),\nn
   & & {1\over 2520}\left({-\frac{35}{N_c+1
   8}+\frac{56}{N_c+21}-\frac{81}{N_c+56}+\frac{35}{N_c+60}+\frac{21}{N_c+630}+
   \frac{4}{N_c}}\right),\nn
   & &{1\over 2520}\left({-\frac{35}{N_c+18}+\frac{56}{N_c+21}-\frac{81}{\text{Nc
   }+56}+\frac{35}{N_c+60}+\frac{21}{N_c+630}+\frac{4}{N_c}}\right)\Biggr\}~~~~~~\eea
%

%%%%%%%%%%%%%%%%%%%%%%%%%%%
%
\bea G^{-1}_{1(41),1(42),1(43),1(44),1(45)} &= & \Biggl\{{1\over 2520}\left(\frac{35}{N_c+18}+\frac{112}{N_c+21}-\frac{117}{N_c+56}-\frac{35}{N_c+60}+\frac
   {21}{N_c+630}-\frac{16}{N_c}\right),\nn
   & & {1\over 2520}\left({-\frac{35}{N_c+18}+\frac{56}{N_c+21}-\frac{81}{N_c+56}+
   \frac{35}{N_c+60}+\frac{21}{N_c+630}+\frac{4}{N_c}}\right),\nn
   & & {1\over 2520}\left({\frac{35}{N_c+18
   }+\frac{28}{N_c+21}+\frac{27}{N_c+56}-\frac{35}{N_c+60}+\frac{21}{N_c+630}
   -\frac{76}{N_c}}\right),\nn
   & & {1\over 2520}\left(\frac{35}{N_c+18}-\frac{56}{N_c+21}-\frac{9}{N_c+56}-\frac{35}{N_c+60}+\frac{21
   }{N_c+630}+\frac{44}{N_c}\right),\nn
   & & {1\over 2520}\left({-\frac{35}{N_c+18}+\frac{56}{N_c+21}-\frac{81}{N_c+56}
   +\frac{35}{N_c+60}+\frac{21}{N_c+630}+\frac{4}{N_c}}\right)\Biggr\}~~~~~~~~~\eea
%

%%%%%%%%%%%%%%%%%%%%%%%%%%%
%
\bea G^{-1}_{1(46),1(47),1(48),1(49),1(50)} &= & \Biggl\{{1\over 2520}\left(\frac{35}{N_c+18}+\frac{112}{N_c+21}-\frac{117}{N_c+56}-\frac{35}{N_c+60}+\frac
   {21}{N_c+630}-\frac{16}{N_c}\right),\nn
   & & {1\over 2520}\left(-\frac{105}{N_c+18}+\frac{45}{N_c+56}-\frac{105}{N_c+60}+\frac{21}{N_c+630}+
   \frac{144}{N_c}\right),\nn
   & & {1\over 2520}\left({\frac{35}{N_c+18}+\frac{28}{N_c+21}+\frac{27}{N_c+56}-\frac
   {35}{N_c+60}+\frac{21}{N_c+630}-\frac{76}{N_c}}\right),\nn
   & & {1\over 2520}\left(\frac{35}{N_c+18}-\frac{56}{N_c+21}-\frac{9}{N_c+56}-\frac{35}{N_c+60}+\frac{21
   }{N_c+630}+\frac{44}{N_c}\right),\nn
   & & {1\over 2520}\left({-\frac{35}{N_c+18}+\frac{56}{N_c+21}-\frac{81}{N_c+56}+
   \frac{35}{N_c+60}+\frac{21}{N_c+630}+\frac{4}{N_c}}\right)\Biggr\}~~~~~~\eea
%

%%%%%%%%%%%%%%%%%%%%%%%%%%%
%
\bea G^{-1}_{1(51),1(52),1(53),1(54),1(55)} &= & \Biggl\{\frac{1}{840}
   \left(\frac{35}{N_c+18}-\frac{56}{N_c+21}-\frac{9}{N_c+56}+\frac{35}{N_c+60}
   +\frac{7}{N_c+630}-\frac{12}{N_c}\right),\nn
   & &  {1\over 2520}\left({-\frac{35}{N_c+18}-\frac{112}{\text
   {Nc}+21}+\frac{27}{N_c+56}+\frac{35}{N_c+60}+\frac{21}{N_c+630}+\frac{64}{\text{Nc
   }}}\right),\nn
   & & {1\over 2520}\left({-\frac{35}{N_c+18}+\frac{56}{N_c+21}-\frac{81}{N_c+56}+\frac{35}{\
   text{Nc}+60}+\frac{21}{N_c+630}+\frac{4}{N_c}}\right),\nn
   & &  {1\over 2520}\left({-\frac{35}{N_c+18}+\frac{56}{N_c+21}-\frac{81}{N_c+56}
   +\frac{35}{N_c+60}+\frac{21}{N_c+630}+\frac{4
   }{N_c}}\right),\nn
   & &  {1\over 2520}\left({\frac{35}{N_c+18}+\frac{28}{N_c+21}+\frac{27}{N_c+56}-
   \frac{35}{N_c+60}+\frac{21}{N_c+630}-\frac{76}{N_c}}\right)\Biggr\}~~~~~~~~\eea
%

%%%%%%%%%%%%%%%%%%%%%%%%%%%
%
\bea G^{-1}_{1(56),1(57),1(58),1(59),1(60)} &= & \Biggl\{ {1\over 2520}\left(\frac{35}{N_c+18}-\frac{56}{N_c+21}-\frac{9}{N_c+56}-\frac{35}{N_c+60}+\frac{21
   }{N_c+630}+\frac{44}{N_c}\right),\nn
   & & {1\over 2520}\left( \frac{35}{N_c+18}-\frac{56}{N_c+21}-\frac{9}{N_c+56}-\frac{35}{N_c+60}+\frac{21
   }{N_c+630}+\frac{44}{N_c}\right),\nn
   & & {1\over 2520}\left({-\frac{35}{N_c+18}-\frac{112}{N_c+21}+\frac{27}{N_c+56}+
   \frac{35}{N_c+60}+\frac{21}{N_c+630}+\frac{64}{N_c}}\right),\nn
   & & {1\over 2520}\left({-\frac{35}{N_c
   +18}-\frac{112}{N_c+21}+\frac{27}{N_c+56}+\frac{35}{N_c+60}+\frac{21}{N_c+63
   0}+\frac{64}{N_c}}\right),\nn
   & &  {1\over 2520}\left({-\frac{35}{N_c+18}+\frac{56}{N_c+21}+\frac{99}{N_c+56}+
   \frac{35}{N_c+60}+\frac{21}{N_c+630}-\frac{176}{N_c}}\right)\Biggr\}~~~~~\eea
   %

%%%%%%%%%%%%%%%%%%%%%
%%%%%%%%%%%%%%%%%%%%%
%%%%%%%%%%%%%%%%%%%%%
%
\subsection*{Remarks}
Before concluding this section, let us make a few remarks on
the degrees of freedom introduced by $N_{c}$.
 First we notice that ${\rm det}(G)$ will depend on $N_c$, thus there are solutions of $N_c$
such that ${\rm det}(G)=0$. When this happens, $G\cdot \tau=n$
will not have solution. In other words,
for these specific values,
 $N_c$ and the imposed loop-KK relations are not compatible to each other. At this moment,
 we are not clear what is
 the physical meaning of these degenerated values of $N_c$.
However from explicit examples discussed above,
 it seems that  $N_c$ that lead to degenerating matrix $G$ are always negative integer.
 For positive $N_c$ there is no problem for it.
It is perhaps possible to choose
special values of $N_c$ such that the final expression  dramatically simplifies or manifest patten
can be observed.

%%%%%%%%%%%%%%%%%%
\section{An alternative approach}
\label{sec:relabeling}
%%%%%%%%%%%%%%%%%%
In previous section, we solve $\tau$ by $n$ using the $G$-matrix directly. Since all conditions we imposed,
such as \eref{map-2}, \eref{tau-double-sym}, \eref{tau-reverse} and \eref{n-tau-final}, are relabeling symmetric,
the solutions $\tau_\sigma$ for different ordering $\sigma$'s are also related by relabeling symmetry.
This property can be used to solve $\tau$ without using the $G$-matrix, which will be the purpose of
 this section. In fact, similar method has been used  in tree-level case in \cite{Du:2013sha}.
For simplicity, in this section we assume $N_{c}=1$.

%%%%%%%%%%%%%%%%%%
\subsection*{Four-point example: }
%%%%%%%%%%%%%%%%%%%

 In the four-point case we  assume that $\tau$ can be expanded by $n$, i.e.,
\bea
\tau_{1234}=an_{1234}+bn_{1243}+cn_{1324}.
\eea
Under the relabeling $1\leftrightarrow 2$, we get
\bea
\tau_{2134}=an_{2134}+bn_{2143}+cn_{2314},
\eea
which can be recast into the original basis
using reflection and cyclic symmetry of $\tau$ and $n$
\bea
\tau_{1342}=an_{1243}+bn_{1234}+cn_{1324}.
\eea
Same $\tau_{1342}$ can also obtained from $\tau_{1234}$ by
relabeling $2\to 3, 3\to 4, 4\to 2$, thus we arrive following equation
\bea
\tau_{1342}=an_{1342}+bn_{1324}+cn_{1432}=an_{1243}+bn_{1324}+cn_{1234}.
\eea
By comparing the $\tau_{1342}$ in this two different ways, we can get
\bea
b=c.
\eea
Thus
\bea
\tau_{abcd}=an_{abcd}+b(n_{abdc}+n_{acbd}).
\eea
Substituting this into the relation between $n$ and $\tau$ \eqref{4-point-rel}, we get
\bea
a={21\over 62},~~~ b=-{5\over 31}.
\eea
Then
\bea
\tau_{abcd}={21\over 62}n_{abcd}-{5\over 31}(n_{abdc}+n_{acbd}).
\eea
This is the same with the result obtained by imposing KK relation and then solving linear equations.

\subsection*{Five-point  expansion }
Similarly at five-points, we assume the dual trace can be %%% following expansion
expanded into the $(5-1)!/2=12$ basis numerators $n_{1,\sigma}$
discussed in section \ref{sec:5-pt-example},
\bea
\tau_{12345} & = & \sum_{\sigma\in S_4/R}c_{1,\sigma}n_{1,\sigma} \\
& = & c_{12345}n_{12345}+c_{12354}n_{12354}+\dots +c_{14325}n_{14325}, \nonumber
\eea
where $R$ denotes reflection.
Comparing the expansion expressions derived through permutating leg $1$ with $2$, $3$, $4$, $5$
with the corresponding expressions obtained by relabeling, we get the following relations
\bea
\tau_{21345}  = -\tau_{12543}  \longrightarrow &c_{12453}=c_{12534}, & c_{12354}=c_{12435}, \\
& c_{13254}=-c_{13425}, & c_{13245}=-c_{14325},\nonumber
 \eea
 \bea
 \tau_{32145}=-\tau_{12354}  \longrightarrow	& c_{12534}=c_{14235}, & c_{12543}=-c_{13245}, \\
& c_{12435}=-c_{14325}, & c_{12453}=-c_{13254}, \nonumber
\eea
 \bea
\tau_{42315}=-\tau_{13245}  \longrightarrow & c_{13254}=-c_{14235}, & c_{12453}=c_{13425} \\
& c_{12435}=-c_{12543}, & c_{12354}=-c_{14325}, \nonumber
 \eea
 \bea
 \tau_{52341}=-\tau_{14325}	  \longrightarrow & c_{12534}=-c_{13254}, & c_{13425}=c_{14235}, \\
& c_{12354}=-c_{12543}, & c_{12435}=c_{13245}.\nonumber
 \eea
Relabling symmetry therefore  reduces the number of independent coefficients to four, yielding
 \bea
 \tau_{12345} & = & a \, n_{12345}+
 + b \, \left(n_{12453}+n_{12534}-n_{13254}+n_{13425}+n_{14235}\right) \\
&& c \, n_{13524} - d \, \left(n_{12354}+n_{12435}-n_{12543}+n_{13245}-n_{14325}\right), \nonumber
 \eea
 while the other basis dual traces $\tau$s can be obtained by relabelings of legs $2$, $3$, $4$ and $5$.
 Substituting these expressions back to just one relation \eref{5-point-rel} allows us to fully determine the
 remaining all four coefficients. Again, we arrive at
\bea & & \tau_{\{1 2 3 4 5\}}  =
{1\over 62}
\left( 16 n_{1 2 3 4 5} -6
n_{1 2 3 5 4}  -6 n_{1 2 4 3 5}
-3 n_{1 2 4 5 3}
-3 n_{1 2 5 3 4}+ 6
n_{1 2 5 4 3}\right.\nn
& &\left. -6 n_{1 3 2 4 5}
+3  n_{1 3 2 5 4}
-3 n_{1 3 4 2 5}-3
n_{1 4 2 3 5} +6 n_{1 4 3 2 5} \right). \eea
%
%%%%%%%%%%%%%%%%%%%%%%%%%%%%%%%%%
\section{Conclusion}
%%%%%%%%%%%%%%%%%%%%%%%%%%%%%%%%%%%
\label{sec:conclusion}
In this work, we have discussed two kinds of dual-color decompositions in Yang-Mills theory at one-loop level. These are the dual-DDM decomposition and the dual-trace decomposition. In both cases, the color-dressed Yang-Mills integrands can be decomposed in terms of color-ordered scalar amplitudes.
We constructed the dual color factors in dual DDM-form by applying Jacobi-like identity for kinematic factors in double-copy formula.
We also constructed the dual-trace factors by imposing KK relation, reflection relation and the relation with the kinematic factor in dual DDM-form.

%%%%%%%%%%%%%%%%%%%%%%%%%%%%%%%%%%%%
\subsection*{Acknowledgements}
%%%%%%%%%%%%%%%%%%%%%%%%%%%%%%%%%%
Y. J. Du would like to acknowledge the EU programme Erasmus Mundus Action 2 and
the International Postdoctoral Exchange Fellowship Program of China for supporting his postdoctoral research in Lund University.
Y. J. Du's research is supported in parts by the NSF of China Grant No.11105118, China Postdoctoral Science Foundation No.2013M530175 and the Fundamental Research Funds for the Central Universities of Fudan University No.20520133169.
C.F. would  like to acknowledge
the support from National Science Council, 50 billions project of
Ministry of Education and National Center for Theoretical Science,
Taiwan, Republic of China as well as the support from S.T. Yau
center of National Chiao Tung University.
B.F is supported, in part,
by fund from Qiu-Shi and Chinese NSF funding under contract
No.11031005, No.11135006, No. 11125523.


\begin{thebibliography}{References}




%\cite{Bern:2008qj}
\bibitem{Bern:2008qj}
  Z.~Bern, J.~J.~M.~Carrasco and H.~Johansson,
  ``New Relations for Gauge-Theory Amplitudes,''
  Phys.\ Rev.\ D {\bf 78}  (2008) 085011
  [arXiv:0805.3993 [hep-ph]].
  %%CITATION = ARXIV:0805.3993;%%


%%%%%%%%%%  Proof of tree-level BCJ


%%%%%%%%%%string proof-KK_BCJ in string theory %%%%%%%%%%%%%%%%%%%%%%%%%%%%%
%\cite{BjerrumBohr:2009rd,Stieberger:2009hq,Tye:2010dd}
\bibitem{BjerrumBohr:2009rd}
  N.~E.~J.~Bjerrum-Bohr, P.~H.~Damgaard and P.~Vanhove,
  ``Minimal Basis for Gauge Theory Amplitudes,''
  Phys.\ Rev.\ Lett.\  {\bf 103} (2009) 161602
  [arXiv:0907.1425 [hep-th]].
  %%CITATION = PRLTA,103,161602;%%
%\cite{Stieberger:2009hq}

\bibitem{Stieberger:2009hq}
  S.~Stieberger,
  ``Open \& Closed vs. Pure Open String Disk Amplitudes,''
  arXiv:0907.2211 [hep-th].
  %%CITATION = ARXIV:0907.2211;%%


%%%%%%%%%%% proof of BCJ form in twistor string %%%%%%%%%%%%%%%%
  %\cite{Cachazo:2012uq, Cachazo:2013gna, Cachazo:2013hca,Cachazo:2013iea }
%\cite{Cachazo:2012uq}
\bibitem{Cachazo:2012uq}
  F.~Cachazo,
  ``Fundamental BCJ Relation in N=4 SYM From The Connected Formulation,''
  arXiv:1206.5970 [hep-th].
  %%CITATION = ARXIV:1206.5970;%%
  %20 citations counted in INSPIRE as of 14 Feb 2014

%\cite{Cachazo:2013gna}
\bibitem{Cachazo:2013gna}
  F.~Cachazo, S.~He and E.~Y.~Yuan,
  ``Scattering Equations and KLT Orthogonality,''
  arXiv:1306.6575 [hep-th].
  %%CITATION = ARXIV:1306.6575;%%
  %15 citations counted in INSPIRE as of 14 Feb 2014

%\cite{Cachazo:2013hca}
\bibitem{Cachazo:2013hca}
  F.~Cachazo, S.~He and E.~Y.~Yuan,
  ``Scattering of Massless Particles in Arbitrary Dimension,''
  arXiv:1307.2199 [hep-th].
  %%CITATION = ARXIV:1307.2199;%%
  %18 citations counted in INSPIRE as of 14 Feb 2014

%\cite{Cachazo:2013iea}
\bibitem{Cachazo:2013iea}
  F.~Cachazo, S.~He and E.~Y.~Yuan,
  ``Scattering of Massless Particles: Scalars, Gluons and Gravitons,''
  arXiv:1309.0885 [hep-th].
  %%CITATION = ARXIV:1309.0885;%%
  %18 citations counted in INSPIRE as of 14 Feb 2014



%%%%%%%%%%%% proof of BCJ in field theory %%%%%%%%%%%%%%%%%%%%%
  %\cite{Feng:2010my, Jia:2010nz, Chen:2011jxa}
   %\cite{Feng:2010my}
\bibitem{Feng:2010my}
  B.~Feng, R.~Huang and Y.~Jia,
  ``Gauge Amplitude Identities by On-shell Recursion Relation in S-matrix Program,''
  Phys.\ Lett.\ B {\bf 695} (2011) 350
  [arXiv:1004.3417 [hep-th]].
  %%CITATION = ARXIV:1004.3417;%%

%\cite{Jia:2010nz}
\bibitem{Jia:2010nz}
  Y.~Jia, R.~Huang and C.~-Y.~Liu,
  ``$U(1)$-decoupling, KK and BCJ relations in $\mathcal{N}=4$ SYM,''
  Phys.\ Rev.\ D {\bf 82}  (2010) 065001
  [arXiv:1005.1821 [hep-th]].
  %%CITATION = ARXIV:1005.1821;%%


%\cite{Chen:2011jxa}
\bibitem{Chen:2011jxa}
  Y.~-X.~Chen, Y.~-J.~Du and B.~Feng,
  ``A Proof of the Explicit Minimal-basis Expansion of Tree Amplitudes in Gauge Field Theory,''  JHEP {\bf 1102} (2011) 112  [arXiv:1101.0009 [hep-th]].  %%CITATION = ARXIV:1101.0009;%%


%%%%%%%%%%%%%%%%%%%%%%%%%%%%%%%%%%%%%%%%%%%%%%%%%%%%%%%%%%%%%%%%%%%%%
%\cite{Sondergaard:2009za,Tye:2010dd, BjerrumBohr:2010zs , Tye:2010kg, Mafra:2011kj, Monteiro:2011pc ,BjerrumBohr:2012mg  ,Fu:2012uy, Monteiro:2013rya}

%\cite{Sondergaard:2009za}
\bibitem{Sondergaard:2009za}
  T.~Sondergaard,
  ``New Relations for Gauge-Theory Amplitudes with Matter,''
  Nucl.\ Phys.\ B {\bf 821}, 417 (2009)
  [arXiv:0903.5453 [hep-th]].
  %%CITATION = ARXIV:0903.5453;%%
  %30 citations counted in INSPIRE as of 14 Feb 2014

  %\cite{Tye:2010dd}
\bibitem{Tye:2010dd}
  S.~H.~Henry Tye and Y.~Zhang,
  ``Dual Identities inside the Gluon and the Graviton Scattering Amplitudes,''
  JHEP {\bf 1006} (2010) 071
  [Erratum-ibid.\  {\bf 1104} (2011) 114]
  [arXiv:1003.1732 [hep-th]].
  %%CITATION = JHEPA,1006,071;%%

%\cite{BjerrumBohr:2010zs}
\bibitem{BjerrumBohr:2010zs}
  N.~E.~J.~Bjerrum-Bohr, P.~H.~Damgaard, T.~Sondergaard and P.~Vanhove,
  ``Monodromy and Jacobi-like Relations for Color-Ordered Amplitudes,''
  JHEP {\bf 1006}, 003 (2010)
  [arXiv:1003.2403 [hep-th]].
  %%CITATION = ARXIV:1003.2403;%%
  %57 citations counted in INSPIRE as of 14 Feb 2014


    \bibitem{Tye:2010kg}
  H.~Tye and Y.~Zhang,
  ``Comment on the Identities of the Gluon Tree Amplitudes,''
  arXiv:1007.0597 [hep-th].
  %%CITATION = ARXIV:1007.0597;%%


%\cite{Mafra:2011kj}
\bibitem{Mafra:2011kj}
  C.~R.~Mafra, O.~Schlotterer, S.~Stieberger and ,
  ``Explicit BCJ Numerators from Pure Spinors,''  JHEP {\bf 1107} (2011) 092  [arXiv:1104.5224 [hep-th]].  %%CITATION = ARXIV:1104.5224;%%  %30 citations counted in INSPIRE as of 30 Mar 2013

%\cite{Monteiro:2011pc}
\bibitem{Monteiro:2011pc}
  R.~Monteiro and D.~O'Connell,
  ``The Kinematic Algebra From the Self-Dual Sector,''
  JHEP {\bf 1107} (2011) 007
  [arXiv:1105.2565 [hep-th]].
  %%CITATION = ARXIV:1105.2565;%%

%\cite{BjerrumBohr:2012mg}
\bibitem{BjerrumBohr:2012mg}
  N.~E.~J.~Bjerrum-Bohr, P.~H.~Damgaard, R.~Monteiro and D.~O'Connell,
  ``Algebras for Amplitudes,''
  JHEP {\bf 1206} (2012) 061
  [arXiv:1203.0944 [hep-th]].
  %%CITATION = ARXIV:1203.0944;%%

%\cite{Fu:2012uy}
\bibitem{Fu:2012uy}
  C.~-H.~Fu, Y.~-J.~Du and B.~Feng,
  ``An algebraic approach to BCJ numerators,''
  JHEP {\bf 1303}, 050 (2013)
  [arXiv:1212.6168 [hep-th]].
  %%CITATION = ARXIV:1212.6168;%%
  %14 citations counted in INSPIRE as of 14 Feb 2014


%\cite{Monteiro:2013rya}
\bibitem{Monteiro:2013rya}
  R.~Monteiro and D.~O'Connell,
  ``The Kinematic Algebras from the Scattering Equations,''
  arXiv:1311.1151 [hep-th].
  %%CITATION = ARXIV:1311.1151;%%
  %7 citations counted in INSPIRE as of 14 Feb 2014

%%%%%%%%%%%%%%%%%%%%%%%%%%%%%

%\cite{Sondergaard:2011iv}
\bibitem{Sondergaard:2011iv}
  T.~Sondergaard,
  ``Perturbative Gravity and Gauge Theory Relations: A Review,''
  Adv.\ High Energy Phys.\  {\bf 2012}, 726030 (2012)
  [arXiv:1106.0033 [hep-th]].
  %%CITATION = ARXIV:1106.0033;%%
  %13 citations counted in INSPIRE as of 14 Feb 2014




%%%%%%%%%%%%%%%%%%%  BCJ-form at the loop level %%%%%%%%%%%%%%%%%%%%
%\cite{BjerrumBohr:2011xe,Boels:2011tp,Boels:2011mn, Du:2012mt,Boels:2012ew, Carrasco:2012ca,Carrasco:2011mn , Bjerrum-Bohr:2013iza, Bern:2013yya, Boels:2013bi, Nohle:2013bfa, Nohle:2013bfa, Bern:2010yg, Du:2011js  }
%\cite{BjerrumBohr:2011xe}
\bibitem{BjerrumBohr:2011xe}
  N.~E.~J.~Bjerrum-Bohr, P.~H.~Damgaard, H.~Johansson and T.~Sondergaard,
  ``Monodromy--like Relations for Finite Loop Amplitudes,''
  JHEP {\bf 1105}, 039 (2011)
  [arXiv:1103.6190 [hep-th]].
  %%CITATION = ARXIV:1103.6190;%%
  %20 citations counted in INSPIRE as of 14 Feb 2014

  %\cite{Boels:2011tp}
\bibitem{Boels:2011tp}
  R.~H.~Boels and R.~S.~Isermann,
  ``New relations for scattering amplitudes in Yang-Mills theory at loop level,''
  Phys.\ Rev.\ D {\bf 85} (2012) 021701
  [arXiv:1109.5888 [hep-th]].
  %%CITATION = ARXIV:1109.5888;%%
  %14 citations counted in INSPIRE as of 18 Jan 2014

  %\cite{Boels:2011mn}
\bibitem{Boels:2011mn}
  R.~H.~Boels and R.~S.~Isermann,
  ``Yang-Mills amplitude relations at loop level from non-adjacent BCFW shifts,''
  JHEP {\bf 1203} (2012) 051
  [arXiv:1110.4462 [hep-th]].
  %%CITATION = ARXIV:1110.4462;%%
  %18 citations counted in INSPIRE as of 18 Jan 2014

  %\cite{Du:2012mt}
\bibitem{Du:2012mt}
  Y.~-J.~Du and H.~Luo,
  ``On General BCJ Relation at One-loop Level in Yang-Mills Theory,''
  JHEP {\bf 1301} (2013) 129
  [arXiv:1207.4549 [hep-th]].
  %%CITATION = ARXIV:1207.4549;%%
  %1 citations counted in INSPIRE as of 18 Jan 2014

%\cite{Boels:2012ew}
\bibitem{Boels:2012ew}
  R.~H.~Boels, B.~A.~Kniehl, O.~V.~Tarasov and G.~Yang,
  ``Color-kinematic Duality for Form Factors,''
  JHEP {\bf 1302} (2013) 063
  [arXiv:1211.7028 [hep-th]].
  %%CITATION = ARXIV:1211.7028;%%
  %19 citations counted in INSPIRE as of 18 Jan 2014

%\cite{Carrasco:2012ca}
\bibitem{Carrasco:2012ca}
  J.~J.~M.~Carrasco, M.~Chiodaroli, M.~G¨¹naydin and R.~Roiban,
  ``One-loop four-point amplitudes in pure and matter-coupled N <= 4 supergravity,''
  JHEP {\bf 1303} (2013) 056
  [arXiv:1212.1146 [hep-th]].
  %%CITATION = ARXIV:1212.1146;%%
  %17 citations counted in INSPIRE as of 18 Jan 2014

  %\cite{Carrasco:2011mn}
\bibitem{Carrasco:2011mn}
  J.~J.~.Carrasco and H.~Johansson,
  ``Five-Point Amplitudes in N=4 Super-Yang-Mills Theory and N=8 Supergravity,''
  Phys.\ Rev.\ D {\bf 85} (2012) 025006
  [arXiv:1106.4711 [hep-th]].
  %%CITATION = ARXIV:1106.4711;%%
  %51 citations counted in INSPIRE as of 18 Jan 2014

%\cite{Bjerrum-Bohr:2013iza}
\bibitem{Bjerrum-Bohr:2013iza}
  N.~E.~J.~Bjerrum-Bohr, T.~Dennen, R.~Monteiro and D.~O'Connell,
  ``Integrand Oxidation and One-Loop Colour-Dual Numerators in N=4 Gauge Theory,''
  JHEP {\bf 1307} (2013) 092
  [arXiv:1303.2913 [hep-th]].
  %%CITATION = ARXIV:1303.2913;%%
  %15 citations counted in INSPIRE as of 18 Jan 2014

%\cite{Bern:2013yya}
\bibitem{Bern:2013yya}
  Z.~Bern, S.~Davies, T.~Dennen, Y.~-t.~Huang, J.~Nohle and ,
  ``Color-Kinematics Duality for Pure Yang-Mills and Gravity at One and Two Loops,''  arXiv:1303.6605 [hep-th].  %%CITATION = ARXIV:1303.6605;%%

 %\cite{Boels:2013bi}
\bibitem{Boels:2013bi}
  R.~H.~Boels, R.~S.~Isermann, R.~Monteiro and D.~O'Connell,
  ``Colour-Kinematics Duality for One-Loop Rational Amplitudes,''
  JHEP {\bf 1304} (2013) 107
  [arXiv:1301.4165 [hep-th]].
  %%CITATION = ARXIV:1301.4165;%%
  %17 citations counted in INSPIRE as of 18 Jan 2014

%\cite{Nohle:2013bfa}
\bibitem{Nohle:2013bfa}
  J.~Nohle,
  ``Color-Kinematics Duality in One-Loop Four-Gluon Amplitudes with Matter,''
  arXiv:1309.7416 [hep-th].
  %%CITATION = ARXIV:1309.7416;%%
  %3 citations counted in INSPIRE as of 18 Jan 2014

%%%%%%%%%%%%%%%%%%%%%%%%%%%%%%%%%


   %\cite{Kleiss:1988ne}
\bibitem{Kleiss:1988ne}
  R.~Kleiss and H.~Kuijf,
  ``MULTI - GLUON CROSS-SECTIONS AND FIVE JET PRODUCTION AT HADRON COLLIDERS,''
  Nucl.\ Phys.\  B {\bf 312} (1989) 616.
  %%CITATION = NUPHA,B312,616;%%



%\cite{Kleiss:1988ne, DelDuca:1999rs}


%%%%%%%%%%%Field  KK BCJ %%%%%%%%%%%%%%%%%%%%%%%%%%%%%%%%%%%%%%%%%%%%%%%%%%%%%%%%%
    %\cite{DelDuca:1999rs}
\bibitem{DelDuca:1999rs}
  V.~Del Duca, L.~J.~Dixon and F.~Maltoni,
  ``New color decompositions for gauge amplitudes at tree and loop level,''
  Nucl.\ Phys.\ B {\bf 571} (2000) 51
  [hep-ph/9910563].
  %%CITATION = HEP-PH/9910563;%%






 \bibitem{KLT} H. Kawai, D. Lewellen and H. Tye, "A Relation Betwwen Tree
Amplitudes of Closed and Open Strings", Nucl.Phys.B269 (1986)1.


%\cite{Du:2011js}
\bibitem{Du:2011js}
  Y.~-J.~Du, B.~Feng and C.~-H.~Fu,
  ``BCJ Relation of Color Scalar Theory and KLT Relation of Gauge Theory,''
  JHEP {\bf 1108} (2011) 129
  [arXiv:1105.3503 [hep-th]].
  %%CITATION = ARXIV:1105.3503;%%


 %\cite{Bern:1999bx}
\bibitem{Bern:1999bx}
  Z.~Bern, A.~De Freitas and H.~L.~Wong,
  ``On the coupling of gravitons to matter,''
  Phys.\ Rev.\ Lett.\  {\bf 84} (2000) 3531
  [arXiv:hep-th/9912033].
  %%CITATION = PRLTA,84,3531;%%


%\cite{Bern:2010yg}
\bibitem{Bern:2010yg}
  Z.~Bern, T.~Dennen, Y.~-t.~Huang and M.~Kiermaier,
  ``Gravity as the Square of Gauge Theory,''
  Phys.\ Rev.\ D {\bf 82}  (2010) 065003
  [arXiv:1004.0693 [hep-th]].
  %%CITATION = ARXIV:1004.0693;%%

%\cite{Bern:2011ia}
\bibitem{Bern:2011ia}
  Z.~Bern and T.~Dennen,
  ``A Color Dual Form for Gauge-Theory Amplitudes,''  Phys.\ Rev.\ Lett.\
   {\bf 107}, 081601 (2011)  [arXiv:1103.0312 [hep-th]].  %%CITATION = ARXIV:1103.0312;%%

%\cite{Bern:2011rj}
\bibitem{Bern:2011rj}
  Z.~Bern, C.~Boucher-Veronneau and H.~Johansson,
  ``N >= 4 Supergravity Amplitudes from Gauge Theory at One Loop,''
  Phys.\ Rev.\ D {\bf 84} (2011) 105035
  [arXiv:1107.1935 [hep-th]].
  %%CITATION = ARXIV:1107.1935;%%
  %39 citations counted in INSPIRE as of 24 Feb 2014

  %\cite{Du:2013sha}
\bibitem{Du:2013sha}
  Y.~-J.~Du, B.~Feng and C.~-H.~Fu,
  ``The Construction of Dual-trace Factor in Yang-Mills Theory,''  arXiv:1304.2978 [hep-th].  %%CITATION = ARXIV:1304.2978;%%

%\cite{Fu:2013qna}
\bibitem{Fu:2013qna}
  C.~-H.~Fu, Y.~-J.~Du and B.~Feng,
  ``Note on Construction of Dual-trace Factor in Yang-Mills Theory,''
  JHEP {\bf 1310} (2013) 069
  [arXiv:1305.2996 [hep-th]].
  %%CITATION = ARXIV:1305.2996;%%
  %2 citations counted in INSPIRE as of 18 Jan 2014

%\cite{Bern:1994zx}
\bibitem{Bern:1994zx}
  Z.~Bern, L.~J.~Dixon, D.~C.~Dunbar and D.~A.~Kosower,
  ``One loop n point gauge theory amplitudes, unitarity and collinear limits,''
  Nucl.\ Phys.\ B {\bf 425} (1994) 217
  [hep-ph/9403226].
  %%CITATION = HEP-PH/9403226;%%
  %770 citations counted in INSPIRE as of 14 Feb 2014











\end{thebibliography}
\end{document}